\documentclass{article}
\usepackage[hyphens]{url}
\usepackage[utf8]{inputenc}
\usepackage{graphicx}
\usepackage[english]{babel}
\usepackage{setspace}
\usepackage{fancyhdr}
\usepackage{array}
\usepackage[margin=1in]{geometry}
\usepackage{float}
\usepackage{amsmath}
\usepackage{amssymb}
\usepackage{bbm}
\usepackage{bm}
\usepackage{scalerel,stackengine}
\usepackage[final]{pdfpages}
\usepackage{longtable}
\usepackage{chngcntr}
\usepackage{placeins}
\usepackage{microtype}
\usepackage{tikz}
\usepackage{makecell}
\usepackage{hyperref}
\usepackage{subcaption}
\usepackage{rotating}
\usepackage{soul}
\usepackage{tabularx} 
\usepackage{pdfpages}
\usepackage[square,numbers]{natbib}
\usepackage{multirow}
\usepackage{booktabs}

\usepackage{authblk}

\hypersetup{
    colorlinks = true,
    citecolor = {blue},
    urlcolor = {blue},
    menucolor = {blue},
    linkcolor = {blue}
}

\makeatletter
\patchcmd{\@maketitle}{\LARGE \@title}{\fontsize{18}{20}\selectfont\textbf{\@title}}{}{}
\makeatother

\title{Optimal allocation strategies in platform trials}  

\author[1]{Marta Bofill Roig}
\author[2]{Ekkehard Glimm}  
\author[3]{Tobias Mielke}
\author[ ]{Martin Posch$^1$ \newline (martin.posch@meduniwien.ac.at) }   

\affil[1]{Section for Medical Statistics, Center for Medical Data Science, Medical University of Vienna, Vienna}
\affil[2]{Advanced Methodology and Data Science, Novartis Pharma AG, Basel} 
\affil[3]{Statistics and Decision Sciences, Janssen-Cilag GmbH}

\date{}         
\begin{document}

\maketitle

\begin{abstract}
Platform trials are randomized clinical trials that allow simultaneous comparison of multiple interventions, usually against a common control. Arms to test experimental interventions may enter and leave the platform  over time. This implies that the number of experimental intervention arms in the trial may change over time.  
	Determining optimal allocation rates to allocate patients to the treatment and control arms  in platform trials is challenging because the change in treatment arms implies that also the optimal allocation rates will change when treatments enter or leave the platform. In addition, the optimal allocation depends on the analysis strategy used. 
	In this paper, we derive optimal treatment allocation rates for platform trials with shared controls, assuming that a stratified estimation and testing procedure based on a regression model, is used to adjust for time trends. We consider both, analysis using concurrent controls only as well as analysis methods based on also non-concurrent controls and assume that the total sample size is fixed.
	The objective function to be minimized is the maximum of the variances of the effect estimators. We show that the optimal solution depends on the entry time of the arms in the trial and, in general, does not correspond to the square root of $k$ allocation rule used in the classical multi-arm trials. We illustrate the optimal allocation and evaluate the power and type 1 error rate compared to trials using one-to-one and square root of $k$ allocations by means of a case study. 
\end{abstract}


\section{Introduction} 

Platform trials compare multiple experimental treatments to a control. They are multi-arm multi-stage trials with the additional feature of allowing arms to enter and leave the trial over time \cite{berry2015platform,Saville2016,Woodcock2017,Meyer2020}.  
As in multi-arm trials, the common trial infrastructure permits shortening the required time and reducing the costs to evaluate new interventions. In addition, the shared control group increases the statistical efficiency compared to separate parallel group trials and requires fewer patients to be allocated to the control group. However, due to the additional flexibility of platform trials, their design and analysis are more complex. 


A major concern when designing and analysing platform trials is the potential presence of time trends, due to, for instance, changes in the patient population being recruited. 
Especially, if the allocation rates between each of the active treatment arms and the control group vary over time, such time trends can lead to biased  treatment effect estimates and hypothesis tests \cite{altman1988hidden,getz2017trial,senn2010hans,julious2008biased}. 
To address such biases, time period-adjusted analyses based on regression models have been proposed \cite{jennison1999group,Simon2011}, where the time periods are defined as the time spans where the allocation ratios stay constant. 

Time trends are of an even larger concern when so-called non-concurrent controls are used for treatment-control comparisons.  Here, for a specific experimental treatment arm,  non-concurrent controls refer to the patients allocated to the control group before the arm under evaluation enters the platform trial. In contrast, concurrent controls are the control group patients randomised concurrently (in time) to those in the treatment arm. 
While including non-concurrent controls in the estimation of treatment effects can increase the power of testing treatment-control differences and reduce the variance of the estimates, they can also introduce bias in the estimates due to time trends if not adjusted for \cite{Sridhara2021a,Lee2021,Saville2022}. Also in this context, time period-adjusted analyses based on regression models have been proposed. They can adjust for potential time trends, and thus control the type 1 error and give unbiased estimates, if time trends in all treatment arms are equal and additive on the model scale \cite{Bofill2021ncc}.  

In this paper, we derive optimal treatment allocation rates for platform trials under the assumption that a time period-adjusted analysis based on regression models is used.  
To understand the principles of optimal allocation strategies in platform trials, we focus on the simple setting of a platform trial with two treatment arms and a shared control, where one of the treatment arms enters when the trial is already ongoing. For this platform trial design, we aim to clarify which design elements the optimal allocation ratios depend on and compare the optimal platform trial design with the optimal multi-arm trial design. For the latter, it is well known that for $k$ experimental treatments (and under some additional assumptions), the standard error of treatment effect estimates is minimized for $1:1:\ldots :1:\sqrt k$ allocation \cite{dunnett1955multiple}. 

Several authors discuss the problem of adding a new treatment arm during the ongoing trial using different optimality criteria and statistical analysis procedures.
\citet{Cohen2015} reviewed statistical methodologies and examples of trials with newly added treatment arms. 
\citet{Choodari-Oskooei2020} centered the attention on the family-wise type 1 error rate when new arms are added. 
\citet{Elm2012} evaluated the operating characteristics of pairwise comparisons of trials adding a new arm over the trial under different approaches. 
\citet{Ren2021} described statistical considerations with respect to type 1 error and power in three-arm umbrella trials. They also discussed the optimal allocation ratio for the control arm in periods in which treatment arms overlap, when minimising the sum of variances of the treatment effect estimators. 
However, to the best of our knowledge, \citet{Bennett2020} is the only article in which the optimal allocation rates in platform trials are investigated. 
They optimised the allocation rates to maximize the probability to find all treatments that are better than control, while assuming that the expected treatment effects were equal for all treatment arms. In their approach, treatment comparisons are based on simple group comparisons with z-tests, using concurrent controls. Concurrent data from different periods (where other treatments may have entered or left the platform and the allocation ratios may have changed) are pooled in this approach. However, in platform trials with time trends and changing allocation ratios over time, this approach can lead to an inflation of the type 1 error rate and biased estimates \cite{Simon2011}. More recently, 
\citet{pan_yuan_ye_2022} addressed the modification of the critical boundaries to control the family-wise error rate and re-estimation of the sample sizes when new arms are added, and, similarly as in \citet{Bennett2020} provided the optimal allocation ratios when minimizing the total sample size to achieve a desirable marginal power level, using only concurrent controls and without adjusting for potential time trends. 

Here, we optimise allocation rates for a testing procedure based on a regression approach, which includes a ``period" effect to account for changing allocation ratios when using period-wise treatment effect estimators and focus the attention on the marginal power. We also consider a different optimisation criterion. Instead of the power to reject all null hypotheses corresponding to effective treatments, we minimise the maximum of the standard errors of the means of the treatment effect estimators resulting from the regression model.  This is asymptotically equivalent to maximizing the minimum power across  treatments, assuming equal treatment effects. This implies (under some regularity assumptions) that, under the optimal design, the power of the different arms will be equal. In particular, for multi-sponsor platform trials, this is a reasonable feature, as all sponsors should get the same chance to demonstrate the efficacy of their treatments in the platform. 
In addition, besides platform trials that only use concurrent data to compare treatments against control, we  consider trials that incorporate non-concurrent controls as well.

The paper is structured as follows. In Section \ref{sec:notation}, we introduce the platform trial designs  considered and introduce the notation. In Section \ref{sec:opt}, we derive optimal allocations for trials using concurrent controls only and, in Section  \ref{sec:opt_ncc}, do so for trials also incorporating non-concurrent controls. 
In Section \ref{sec:casestudy}, we illustrate the application of different allocation rules (optimal compared to equal allocation and square root of $k$ allocation) by a simulation study based on an example in the context of a hypercholesterolemia trial.
We conclude the paper with a discussion.

\section{Trial designs and optimality criterion}\label{sec:notation}
Consider a platform trial evaluating the efficacy of two experimental arms ($i=1,2$) against a shared control ($i=0$). 
The trial design allows the sequential entry and exit of the experimental arms, such that the trial initially starts with arm 1 and the control, and arm 2 may enter the platform trial at a later time point. In addition, recruitment to arm 1 may end before the recruitment to arm 2 ends. 
Thus, the platform trial  consists of three periods  ($s=1,2,3$) defined according to the sets of actively recruiting treatment arms $I_s$. In period 1, patients are recruited to treatment 1 and control ($I_1=\{0,1\}$), in period 2 to both experimental treatments and control ($I_2=\{0,1,2\}$) and, in period 3, to  treatment 2 and control ($I_3=\{0,2\}$).  

\begin{figure}[h!]
	\centering
	\includegraphics[width=0.8\linewidth]{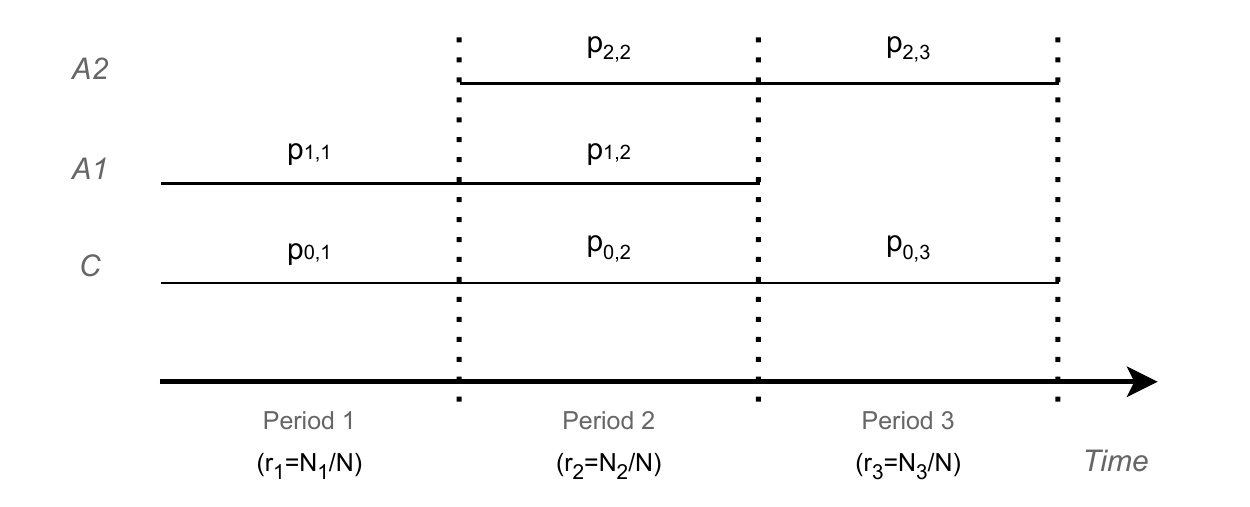}
	\caption{Platform trial with two active treatments and three periods. $p_{i,s}$ denotes the allocation proportions of patients in period $s$ ($s=1,2,3$) to arm $i$ ($i=0,1,2$), $r_s=N_s/N$ refers to the proportion of patients in period $s$, where $N_s$ and $N$ are the sample sizes in period $s$ and in the overall trial, respectively.}
	\label{fig:3armspaper}
\end{figure}

The total sample size of the trial is denoted by $N$, which is partitioned into the three periods with sample sizes $N_s$, $s=1,2,3$. We refer to the  corresponding proportions of patients by $r_s=N_s/N$. In each period $s$,  patients are allocated to the arms $i\in I_s$ with the allocation proportions $p_{i,s}$, such that  $\sum_{i\in I_s} p_{i,s}=1$ for $s=1,2,3$. See Figure \ref{fig:3armspaper} for an illustration of the trial design.

Let $y_{j}$ denote the observation of patient $j$ on treatment arm $i_j$ ($j=1, ..., N$, $i_j=0,1,2$), distributed as $y_{j}\sim N(\mu_{i},\sigma^2)$ where $\sigma$ is assumed to be known and where we dropped the subindex $j$ in the mean for simplicity.
Let $\theta_i=\mu_i-\mu_0$ denote the treatment effect for treatment $i$ ($i=1,2$), and consider the null hypotheses $H_i: \theta_i\leq 0$. 

The simple mean difference between the treatment and control groups are biased estimators of the treatment effects if there are time trends (i.e., the means in the treatment arms change over time)  and the allocation rates change across periods. Therefore, we use stratified estimators, stratified by period, that adjust for potential time trends \cite{Jennison1999,senn2010hans,senn2000many}. The stratified estimators  are given by 
\begin{eqnarray}\label{eff}
	\widehat{\theta}_i &=& \sum_{s=i, i+1} w_{i,s}\cdot \hat{\theta}_{i,s}
\end{eqnarray}
where $\hat{\theta}_{i,s}=\bar{y}_{i,s}-\bar{y}_{0,s}$ 
is the treatment effect estimate in period $s$ for arm $i$ ($i=1,2$), where  $\bar{y}_{i,s}$ and  $\bar{y}_{0,s}$ are the sample means in period $s$ for experimental and control arms, and the weights are given by  
\begin{eqnarray}\label{weight_theta}
	w_{1,s} = \frac{\frac{1}{\sigma_{1,s}^2}}{\frac{1}{\sigma_{1,1}^2} + \frac{1}{\sigma_{1,2}^2}} & \text{and} &
	w_{2,s} = \frac{\frac{1}{\sigma_{2,s}^2}}{\frac{1}{\sigma_{2,2}^2} + \frac{1}{\sigma_{2,3}^2}},
\end{eqnarray}
where  $\sigma_{i,s}^2=\mathrm{Var}(\hat{\theta}_{i,s})$ denotes the variances of the estimates per period, and is given by 
$\sigma_{i,s}^2 =\sigma^2 (1/n_{i,s}+1/n_{0,s})$. This choice of weights minimizes the variance of the stratified treatment effect estimator. 
Then, the variance of the stratified effect estimator is given by: 
\begin{eqnarray} 
	\mathrm{Var}(\widehat{\theta}_i)  
	&=& 
	\sum_{s=i,i+1} w_{i,s}^2 \sigma_{i,s}^2  
	= 
	\frac{\sigma^2}{N} \cdot \left(r_i\frac{p_{i,i}p_{0,i}}{p_{i,i}+p_{0,i}} +r_{i+1} \frac{p_{i,i+1}p_{0,i+1}}{p_{i,i+1}+p_{0,i+1}} \right)^{-1}. \label{var}
\end{eqnarray}  

A few comments: (i) The stratified estimator is equivalent to the non-stratified estimator if the allocation proportions are equal across periods. (ii) The expressions \eqref{var} for the variances apply if the above distributional assumptions hold, i.e, there are  no time trends. However, as discussed above, the stratified treatment effect estimate used is also unbiased if time trends are present. 
(iii) The stratified estimator (and corresponding test) corresponds to the treatment effect estimators in the linear models:
\begin{align}  \label{model_cc_t1} 
	E(y_{j})&= \eta_1 + \theta_1 \cdot I(i_j = 1) + \theta_2 \cdot I(i_j = 2) + \tau_2 \cdot I(j \in (N_1+1,N_2])\\
	E(y_{j})&= \eta_2 + \theta_1 \cdot I(i_j = 1) + \theta_2 \cdot I(i_j = 2) +  \tau_3 \cdot I(j \in (N_2+1,N])  \label{model_cc_t2} 
\end{align} 
where the first model (to test $H_1$) is fit with observations from periods 1 and 2 of the control and treatment arms 1 and 2 and the second model (to test $H_2$) with observations from periods 2 and 3 of the control and arms 1 and 2. Here  $\tau_s$ denotes the period effect which adjusts for  potential time trends. The treatment effects are assumed to be constant in time (no time-by-treatment interaction).

The optimality criterion we aim to minimise is the maximum of the variances of the stratified effect estimators across experimental treatment arms. Thus, given a fixed overall sample size $N$, we aim to find the  allocation probabilities $\mathbf{p}=(p_{0,1},p_{1,1},p_{0,2},p_{1,2},p_{2,2},p_{0,3},p_{2,3})$ that minimise the objective function
\begin{equation}
	\max(\mathrm{Var}(\hat{\theta}_1), \mathrm{Var}(\hat{\theta}_2)). \label{eq:objecive}
\end{equation}
Note that if the expected effect sizes are equal for both experimental arms, minimizing this objective function will also maximize the minimum individual power. Furthermore, as  $N$ is only a scaling factor in (\ref{var}), the optimal allocation does not depend on $N$.


\section{Optimal Designs}\label{sec:opt}

We derive optimal designs minimizing the objective function (\ref{eq:objecive}) for three different cases: 
\begin{description}
	\item[\textbf{Case 1: Unrestricted optimization.}] In this setting, the entry time for treatment 2 (corresponding to $r_1$) is a design parameter that is determined by the trial design rather than being externally governed, e.g., by the time a new treatment becomes available to be included in the trial. Similarly, the time from entry of treatment 2 and completion of the sub-trial corresponding to treatment 1,  $r_2$, is a design parameter in this case, which can also be chosen. We therefore optimize over $\mathbf{p}$ and $r_1,r_2$ ($r_3$ is then given by $r_3=1-r_1-r_2$).
	\item[\textbf{Case 2: Fixed sample size in period 1.}] We fix $r_1$ and optimize $\mathbf{p}$ and $r_2$. In this case, the entry time of arm 2 is not subject to optimization.
	\item[\textbf{Case 3: Fixed sample sizes in all three periods.}] In addition to $r_1$, we also fix $r_2$ (and thus also $r_3$). Therefore, this case corresponds to a design in which both $r_1$, the entry time of arm 2 and the completion time of arm 2 (given by $r_1+r_2$) are predefined, and we optimize over $\mathbf{p}$ only.  
\end{description}

In the derivations, we assume that the sample sizes are positive real numbers such that the variance of the estimators \eqref{var} is a differentiable function of the sample sizes. This is a reasonable approximation for large sample sizes. The derivations were performed in Mathematica (see the Supplementary Material, Section C and GitHub (\url{https://github.com/MartaBofillRoig/Allocation}). 
In calculations resulting in several solutions, we selected those resulting in real values between 0 and 1 based on numeric examples only.

Furthermore, since stratified test statistics are used, the optimal allocation in periods 1 and 3 is equal allocation, i.e., $p_{0,1}=p_{1,1}=p_{0,3}=p_{2,3}=1/2$. This can be directly seen from equation \eqref{var}:  the allocation ratios in period 1 only affect $Var(\hat \theta_1)$ and, since $p_{0,1}=1-p_{1,1}$, it follows that $p_{1,1}p_{0,1}/(p_{1,1}+p_{0,1})$ is maximized for $p_{1,1}=p_{0,1}=1/2$. A similar argument holds for the optimal allocation in period 3. Thus, it remains to determine the optimal allocation in period 2 and, depending on the case  considered, the optimal partition between periods by means of $r_s$.

\subsection{Case 1: Unrestricted optimization}\label{sec:case1}
In this case, the optimal design satisfies $r_2=1$ (and $r_1=r_3=0$) such that all patients are recruited in period 2. In this design, there is only a single period and the effect estimates reduce to the non-stratified mean differences. The one-period design is optimal because then all control observations are shared and are used in both treatment effect estimates, and a non-stratified estimate is applied which, under the model assumption of no time trend, also reduces variance.  The resulting trial is a classical multi-armed trial with many-to-one comparisons \cite{dunnett1955multiple}. For this design, it is well known that the optimal allocation is $p_{0,2}=1/(1+\sqrt{2})$, $p_{1,2}=p_{2,2} =1-1/\sqrt{2}$.

Note that the optimal allocation satisfies the constraint
\begin{equation}
	Var(\widehat \theta_1)=Var(\widehat \theta_2)\label{eq:eqvar}.
\end{equation}
This follows by the following argument: Assume the variances at the optimal allocation are not equal, e.g., that $Var(\widehat \theta_1)<Var(\widehat \theta_2)$. Then the objective function can be further decreased by moving a small fraction of the sample size from treatment 1 to treatment 2 such that the inequality still holds. This shift reduces the $Var(\widehat \theta_2)$ and thereby the maximum of the two variances. This is a contradiction to the assumption that the allocation is optimal.

\subsection{Case 2: Fixed sample size in period 1\label{sec:case2}}

Assume that the time point when the second treatment enters the platform trial is given. Thus, the size of  period 1, $r_1$, is fixed and we optimize (\ref{eq:objecive}) in $ r_2$ and $\mathbf{p}$. As noted above, the optimal allocation in the first and third periods is equal allocation. For the optimal allocation in period 2, we distinguish between the cases $r_1>1/2$, where period 1 is larger than half of the total sample size and therefore for all allocations the variability of $\theta_2$ is larger than that of $\theta_1$  and $r_1\leq 1/2$ where under the optimal allocation the variances of the two estimates are equal.

If $r_{ 1}>1/2$, then, for all allocations (even if all observations after arm 2 enters are allocated to arm 2 and the control), we have $Var(\hat \theta_2)>Var(\hat \theta_1)$ such that in this case also the optimum allocation does not satisfy \eqref{eq:eqvar}. However,  as the objective function is the maximum of the two variances, it  is minimized in this case, if all patients $j>N_1$ are equally allocated to treatment 2 and control, i.e., for $p_{0,1}=p_{1,1}=1/2$, $p_{0,3}=p_{2,3}=1/2$, 
and $r_2=0$. This optimal solution corresponds to two separate consecutive trials, wherein the first period we compare arm 1 versus control, there is no second period, and in the third period, we  use the remaining sample size to compare arm 2 versus control. Note that all design settings comprising only two periods and two arms per period, that is $p_{0,2}=p_{2,2}=1/2$, $p_{1,2}=0$ with $0\leq  r_2\leq 1-r_1$ and $r_3=1-r_1-r_2$, minimize the variance and would be optimal solutions too. In these designs, the allocation ratio is the same in periods 2 and 3, and all patients in periods 2 (and 3) are equally allocated to treatment 2 and control.

Otherwise, if $r_{1}\leq 1/2$,  similar as in Case 1 the optimal design satisfies that the variances of the effect estimators are equal \eqref{eq:eqvar}. Furthermore, for the optimal design we have $r_2=1-r_{1}$ and $r_3=0$. While this can be confirmed by numerical optimisation -- see Figure \ref{fig:optiplat} where the maximum of the two variances compared to separate trials is plotted with respect to $r_2$ and note that the maximum variance reduction occurs for $r_1 + r_2=1$, there is also a heuristic argument for this result: For any given three-period platform trial (with $r_1,r_2,r_3>0$), we can define a corresponding two-period platform trial with the same sample sizes in each arm. This can be obtained by shifting the observations from arms $i=0,2$ in period 3 to the respective arms in period 2 such that $r_3=0$. Now, by recruiting all control arm patients from period 2 onward in period 2, instead of splitting them between periods 2 and 3,  the sample size of the control arm in the estimate of $\widehat\theta_1$ increases and therefore the variance of the estimator is reduced. In addition, the variance of $\widehat\theta_2$ is not increased in this design, as the number of observations in the control arm used in the estimate of $\widehat\theta_1$  does not change and, as $r_3=0$, a non-stratified estimate is used, which has a lower variability than the stratified estimate (under the model assumption of no time trend).

The optimal allocations $p_{0,2}$, $p_{1,2}$ and $p_{2,2}$ can be obtained numerically as special case of Case 3, setting $r_2=1-r_1$, outlined below. Hence, the optimal design is a two-period trial in which arm 2 enters later, but both arms finish at the end of the trial.

\subsection{Case 3: Fixed sample sizes in all three periods\label{sec:case3}}

Assume that, in addition to  the time point when treatment 2 enters, $r_{1}$, also  the number of patients in period 2, $r_2$, is fixed. 
For the optimal allocation in period 2, we now consider three situations.
(i) If $r_1\geq 1/2$, as in Case 2, the optimal allocation allocates all patients after period 1 to treatment arm 2 and control, with equal allocation between the two arms. This is achieved, e.g., for $p_{0,1}=p_{1,1}=p_{0,2}=p_{2,2}=p_{0,3}=p_{2,3}=1/2$ and $p_{1,2}=0$. (ii) If $r_{1}+r_{2}\leq 1/2$, then for all allocations, the maximum variance is the variance of $\hat \theta_1$ and therefore (as in Case 2) the optimal design allocates all observations in period 2 to treatment 1, such that  $p_{0,2}=p_{1,2}=1/2$, $p_{2,2}=0$. (iii) If, on the other hand, $r_1<1/2$ and $r_{1}+r_{2}>1/2$, the argument as for Case 2 implies that the variances of the two treatment effect estimators are equal,  \eqref{eq:eqvar}, holds under the optimal allocation.
To optimise under this constraint, we use the method of Lagrange multipliers, which shows that the solution satisfies
\begin{eqnarray}
	p_{0,2}&=& \frac{1}{2 (1-p_{2,2})} - p_{2,2}\label{p21_sol}\\
	\frac{r_2}{1-2 r_1}&=& 
	\frac{(1-p_{2,2})^3 }{(2 p_{2,2}-1) (p_{2,2} (p_{2,2} (p_{2,2} (2 p_{2,2} (2 p_{2,2}-7)+19)-15)+7)-2)}\label{p22_sol}
\end{eqnarray}

Now, $p_{2,2}$ can be obtained as the numerical solution of \eqref{p22_sol} and  $p_{0,2}$ is given by \eqref{p21_sol}. As the sum of the allocation proportions is 1 also $p_{1,2}$ results.  Note that by \eqref{p22_sol} that the optimal solution depends on $r_1,r_2$ only via $r_2/(1-2r_1)$.

\begin{figure}[h!]
	\centering
	\begin{tabular}{cc}
		\includegraphics[width=0.45\textwidth]{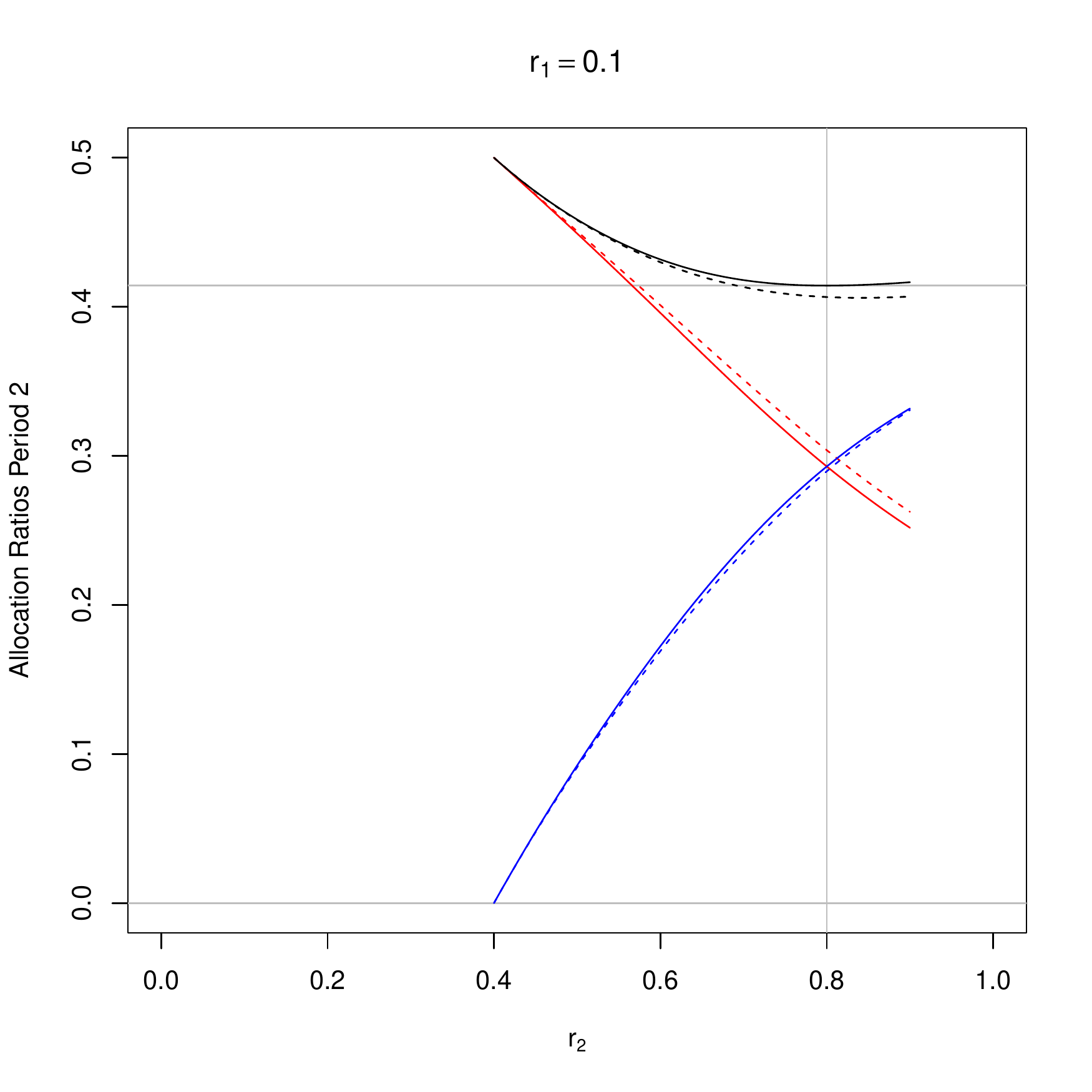}&
		\includegraphics[width=0.45\textwidth]{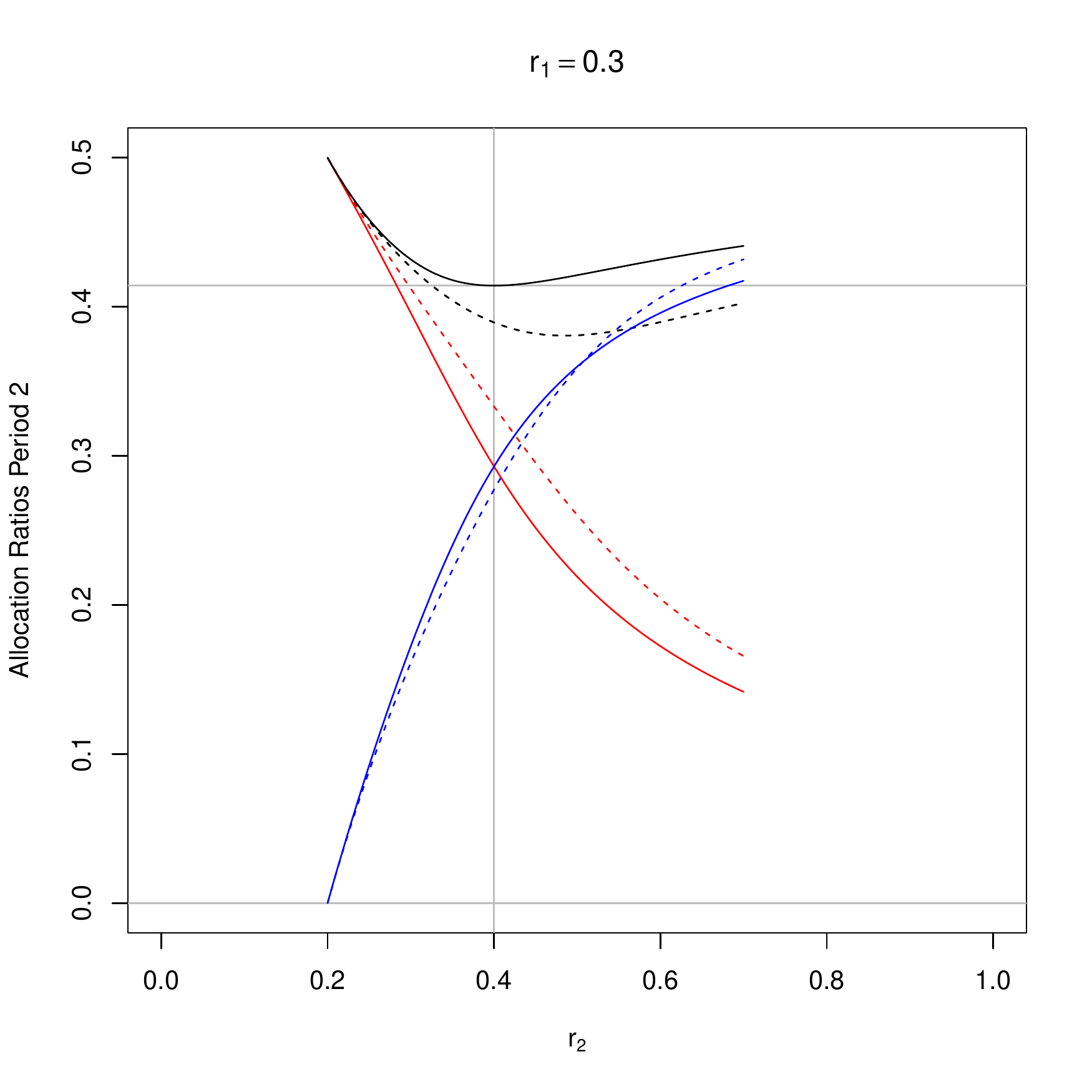}\\
		\includegraphics[width=0.45\textwidth]{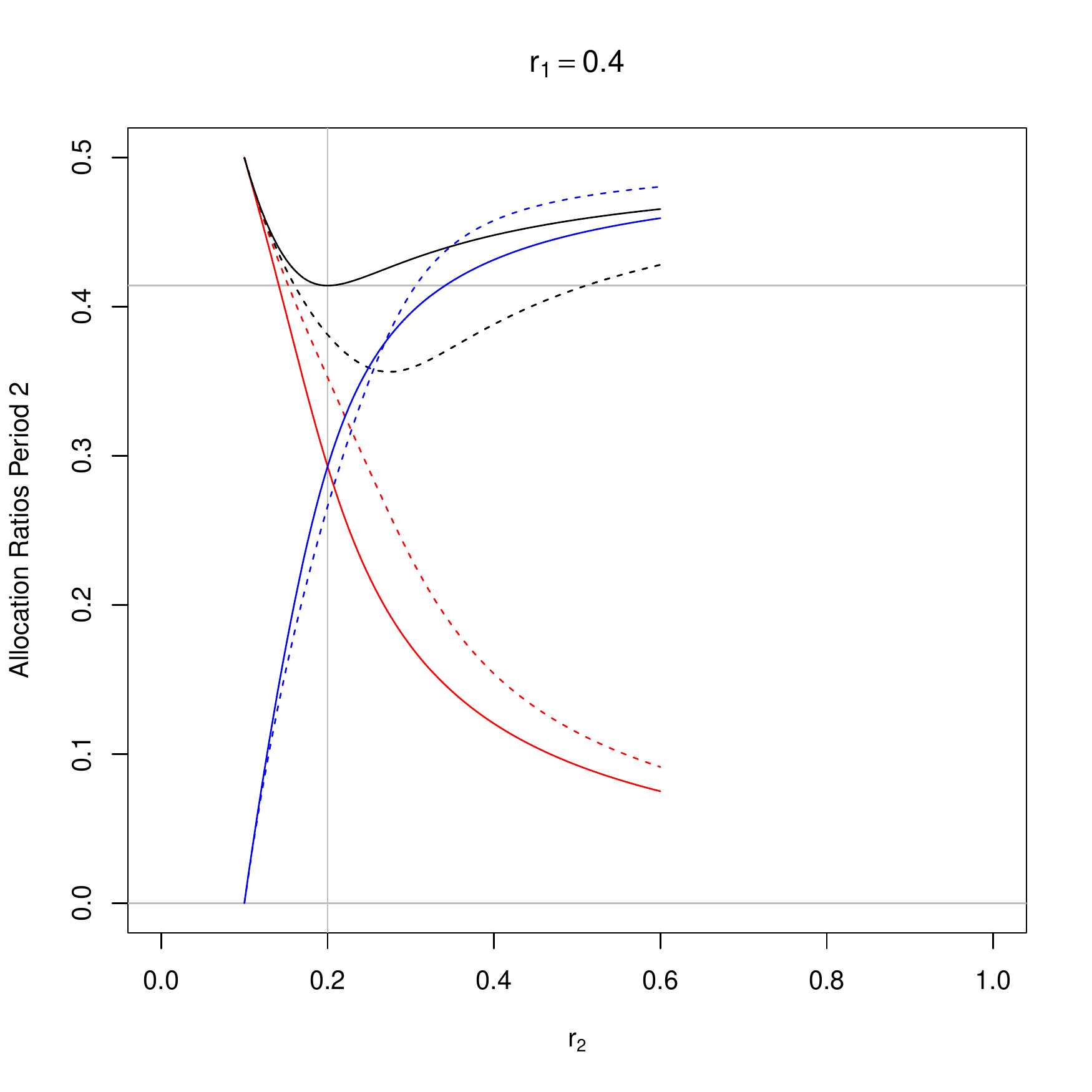}&
		\includegraphics[width=0.45\textwidth]{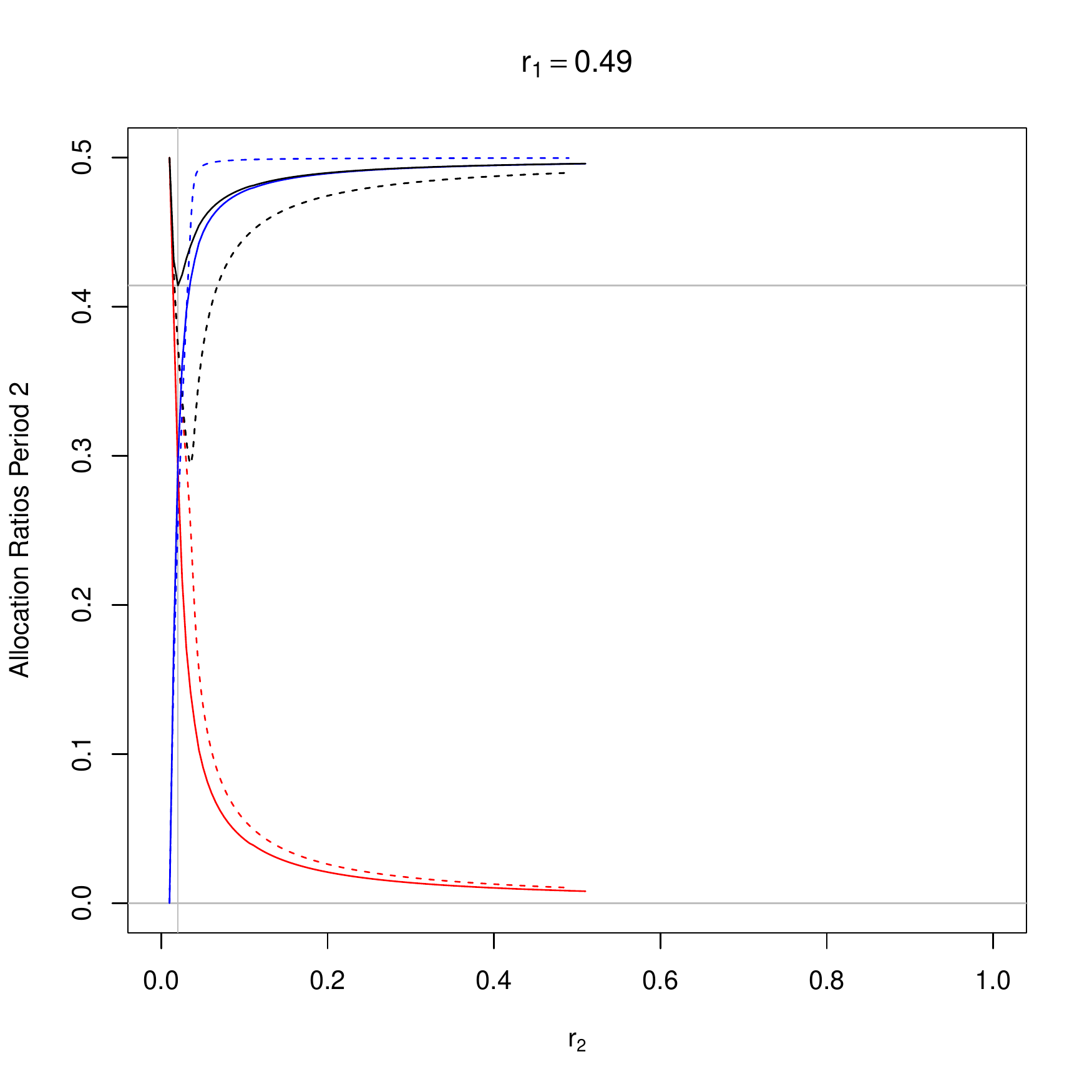}
	\end{tabular}
	\caption{Optimal allocation probabilities $p_{i,2}, i=0,1,2$ in period 2 as function of $r_2$ and different $r_1$ for trials with three-periods as in Figure \ref{fig:3armspaper}. Black lines: $p_{0,2}$, red line $p_{1,2}$, blue line $p_{2,2}$. The vertical line at $r_2=1-2*r_1$ indicates the setting, where equal allocation between treatments 1 and 2 is optimal. There the optimal allocation ratio to control is $p_{0,2}=1/(1+\sqrt{2})$ (indicated by the upper gray line). Solid lines correspond to optimal allocations in trials utilising concurrent controls only (Sect. \ref{sec:opt}), and dashed lines correspond to optimal allocations in trials utilising concurrent and non-concurrent controls (Sect. \ref{sec:opt_ncc}).}
	\label{fig:case3}
\end{figure}

Figure \ref{fig:case3} shows the optimal allocation probabilities $p_{i,2},i=0,1,2$ in period 2 as function of $r_2$ for different $r_1$.  
The optimal allocation ratio $p_{0,2}$ for the control is not monotone in $r_2$ but 
takes its minimum at $r_2=1-2 r_1$, i.e., where $r_1=r_3$. For this special case, we can derive an explicit solution for the optimisation problem. As in this case, the objective function is symmetric in the two treatments, it follows that $Var(\hat \theta_{1,1})=Var(\hat \theta_{2,3})$. Therefore, to satisfy constraint \eqref{eq:eqvar}, also $Var(\hat \theta_{1,2})=Var(\hat \theta_{2,2})$ needs to hold. The latter, however, implies that in period 2 the allocation ratios for both treatment arms have to be equal, such that $p_{1,2}=p_{2,2}$. Furthermore, by results for the many-to-one setting \cite{dunnett1955multiple}, we know that the  variances are minimized for $1:1:\sqrt{2}$ allocation, such that $p_{0,2}=1/(1+\sqrt{2})$ and $p_{1,2}=p_{2,2}=1-1/\sqrt{2}$. This can be also verified by substituting the solution in \eqref{p22_sol}.

\begin{figure}[h!]
	\centering
	\includegraphics[width=1\linewidth]{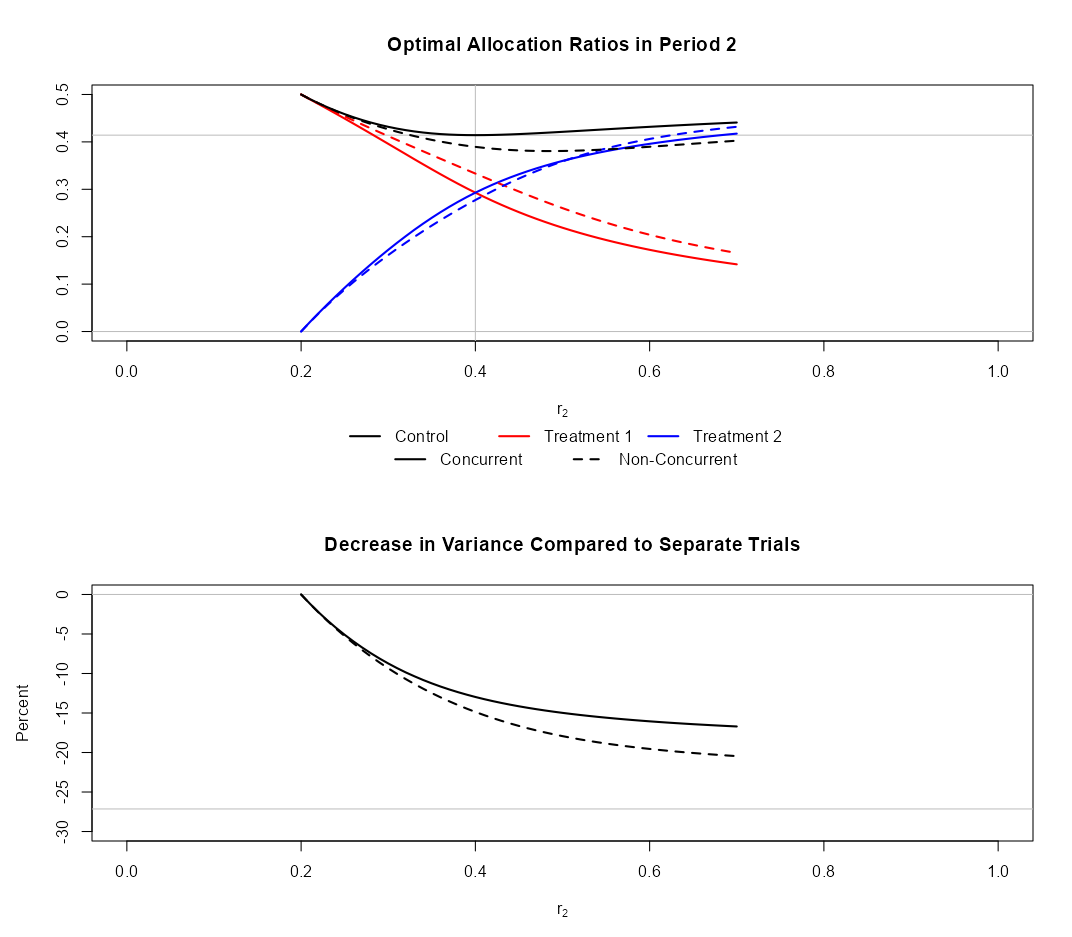}
	\caption{Optimal allocations per arm in period 2 and decrease in the maximum of the two variances compared to separate trials for trial designs using concurrent controls only (solid line) and using concurrent and non-concurrent controls (dashed line). }
	\label{fig:optiplat}
\end{figure}

In order to illustrate further cases apart from those included in the paper, we built a shiny app (available at \url{https://github.com/MartaBofillRoig/Allocation}) that allows visualising the allocation rates with respect to the sample sizes per period 2. The app also shows the variance in the optimum as a function of $r_2$. Figure \ref{fig:optiplat} includes one of the figures that can be obtained in the app. On the bottom, we see the variance  with respect to the proportion of $r_2=N_2/N$ compared to the variance when running two separate trials. We can observe that the decrease in variance becomes larger with $r_2$. 

\section{Optimal allocation in trials using non-concurrent controls} \label{sec:opt_ncc} 

Suppose we want to use also non-concurrent controls to compare arm 2 against the control. To this end, we consider the regression model estimate of the treatment effect of treatment 2 from the model
\begin{align}  \label{model_cc} 
	E(y_{j})&= \eta_0 + \sum_{i=1,2} \theta_i \cdot I(i_j = i) + \sum_{s=2,3} \tau_s \cdot I(j \in (N_{s-1}+1,N_s]),
\end{align} 
We fit this model based on all observed data in periods 1, 2 and 3. This model provides unbiased treatment effect estimators and controls the type 1 error if time trends affect all arms equally and are additive on the model scale \cite{Bofill2021ncc}. For treatment 1, we suppose that the efficacy is evaluated after treatment arm 1 ends, as is common in platform trials. Thus, we do not use the estimate from this model, but the stratified treatment effect estimators defined in \eqref{model_cc_t1} based on data from treatment 1 and the  control group  from stages 1 and 2 only.

Using the regression model \eqref{model_cc} to incorporate non-concurrent controls implies that the treatment effect of arm 2 compared to control is estimated by the  stratified treatment effect estimator 
\begin{eqnarray}\label{effect2_ncc}
	\widetilde\theta_2 &=&
	\omega_{1,1} (\bar{y}_{1,1}-\bar{y}_{.,1})  + 
	\omega_{1,2} (\bar{y}_{1,2}-\bar{y}_{.,2}) +
	\omega_{2,2} (\bar{y}_{2,2}-\bar{y}_{.,2}) +
	\omega_{2,3} (\bar{y}_{2,3}-\bar{y}_{.,3}), 
\end{eqnarray}
where $\bar{y}_{.,s}$ is the pooled mean per period, and $\omega_{i,s}$ are weights, which are functions of the sample sizes. 
As a consequence, also the variance of $\widetilde\theta_2$  can be written as a function of the rates $\mathbf{r}$ and the total sample size $N$. This variance is given by 
\begin{equation}\label{vardelta2inv}
	\mathrm{Var}(\widetilde\theta_2)=\frac{\sigma^2}{N} \left( r_3 p_{2,3}(1-p_{2,3})+r_2\left( p_{2,2}(1-p_{2,2})-\frac{r_2 p_{1,2}^2 p_{2,2}^2}{r_1 p_{1,1}(1-p_{1,1})+r_2 p_{1,2}(1-p_{1,2})}\right)\right)^{-1}.
\end{equation}
See the Supplementary Material Section A2 for the derivation and the expression of the weights $\omega_{i,s}$. For arm 1, for which the treatment effect is estimated using only concurrent controls, the treatment effect estimator and its variance are given by \eqref{eff} and \eqref{var}, respectively. 

As in Section \ref{sec:opt}, we aim to minimise the maximum of the variances $\max(\mathrm{Var}(\widehat{\theta}_1),\mathrm{Var}(\widetilde\theta_2))$ in  $\mathbf{r}$ and discuss the optimal allocations for unrestricted optimisation (Case 1), where the entry time of treatment 2 is fixed (Case 2) and where both the entry time of treatment 2 and the completion time of treatment 1 are given (Case 3). 

Note that it is evident from \eqref{var} and \eqref{vardelta2inv} that for all $r_1,r_2,r_3$ (and thus for all three considered cases), the optimal allocation in periods 1 and 3 is equal allocation, i.e., $p_{0,1}=p_{1,1}=p_{0,3}=p_{2,3}=1/2$. Therefore, it remains to determine the optimal allocation ratios in period 2 in the three cases. 

\subsection{Case 1: Unrestricted optimization}
As for the setting using concurrent controls only,  without  restrictions with respect to the time at which arm 2 enters the platform, the optimal allocation rates correspond to a multi-arm design where all treatment arms start and end at the same time, such that there are no non-concurrent controls. This follows because then all control patients are used in direct comparisons for both treatment arms. Then, the optimal allocation is  $1:1:\sqrt{2}$ allocation as in Section \ref{sec:opt}.  

\subsection{Case 2: Fixed sample size in period 1}
Suppose we aim at optimising a design where the time at which arm 2 enters is given. Therefore, as in Case 2 for trials using concurrent controls only (Section \ref{sec:case2}), we optimise assuming that $r_1$ is fixed. 

We  consider two cases: $r_1\geq 1/2$ and $r_1<1/2$. 
If $r_1\geq 1/2$, as in the case of concurrent controls, the optimal allocation allocates all patients after period 1 to treatment arm 2 and control, with equal allocation between the two arms. This is achieved, e.g., for $p_{0,1}=p_{1,1}=1/2$,
$p_{0,3}=p_{2,3}=1/2$ and $p_{i,2}=0$, $i=0,1,2$ such that $r_2=0$. 
Note that this design has only two periods and the model \eqref{model_cc} is fitted without the period effect $\tau_2$ as there is no data to estimate this factor in this case. In addition, the non-concurrent controls do not contribute to the treatment effect estimator (only the weight $\omega_{2,3}$ in  \eqref{effect2_ncc} is larger than zero) because only the control treatment is present in both periods. Therefore, the period effect can only be estimated from the control group data and consequently the  estimate of the control group treatment (and also the treatment effect estimate for treatment 2) cannot be improved by estimates from other treatments.

To see that the above allocation is optimal, note that  $Var(\tilde \theta_2)$ defined in \eqref{vardelta2inv} 
is minimized by $r_3=1-r_1,r_2=0$ and $p_{0,3}=p_{2,3}=1/2$ if $r_1$ is kept fixed and for this allocation the variance  $Var(\tilde \theta_2)$ does not depend on the allocation ratios in period 1. Furthermore, because $r_1\geq 1/2$ and assuming equal allocation between the control arm and arm 1 in period 1 we have $Var(\hat \theta_1)\leq Var(\tilde \theta_2)$. Therefore, under an allocation ratio that minimizes $Var(\tilde \theta_2)$ the objective function, the maximum of the variances of the treatment effect estimates is the variance of treatment 2. It follows, that this allocation ratio also minimizes the objective function.    

If $r_1<1/2$, as in Case 2 in Section \ref{sec:opt}, the optimal design satisfies that $r_2=1-r_1$, and therefore it leads to a two-period platform trial, where treatments 1 and 2 complete recruitment at the same time and there is no period 3. 
While this can be seen by inspection of Figure \ref{fig:optiplat}, there is also an argument for this result. First, note that for any given three-period platform trial (with $r_2,r_3>0$), we can achieve the same sample sizes for each arm in a two-period platform trial, shifting all observations from period 3 to period 2 such that $r_3=0$. Now, after period 1, recruiting all observations of the control arm in period 2 instead of splitting them between periods 2 and 3, increases the sample size of the control arm for the estimate of $\widehat\theta_1$ and therefore reduces its variance. Also, the variance of $\tilde\theta_2$ is not increased, as the number of concurrent controls for arm 2 is not affected and, as $r_3=0$, the model \eqref{model_cc} is fitted with one parameter less (without $\tau_3$) in this case.

It remains to determine the optimal allocation in period 2 for the resulting two-period platform trial. In this case, the effect size estimate of treatment 2  using non-concurrent controls  \eqref{effect2_ncc} can be written as 
\begin{eqnarray}\label{theta22_ncc}
	\widetilde\theta_2 &=&
	\hat \theta_{2,2} + \rho (\hat \theta_{1,1}-\hat \theta_{1,2})  
\end{eqnarray} 
where
$   \rho = { n_{0,2}^{-1} }/( n_{0,1}^{-1} +  n_{0,2}^{-1} +n_{1,1}^{-1} + n_{1,2}^{-1})$
and its variance simplifies to: 
\begin{eqnarray}\label{var_ncc}
	\mathrm{Var}(\widetilde\theta_2) &=& 
	\frac{\sigma^2}{N}\left( r_2 q_{2,2}-\frac{r_2^2 p_{1,2}^2 p_{2,2}^2}{r_1 q_{1,1}+r_2 q_{1,2}} \right)^{-1}, \label{var_ncc2}
\end{eqnarray}
where  $q_{i,s}=p_{i,s}(1-p_{i,s})$. 

We now minimise 
$\max(\mathrm{Var}(\widehat{\theta}_1),\mathrm{Var}(\widetilde\theta_2))$, where
$\mathrm{Var}(\widehat{\theta}_1)$ and $\mathrm{Var}(\widetilde\theta_2)$
are defined by \eqref{var} and \eqref{var_ncc}, respectively.
For given $r_1$ and assuming that  $r_3=0$ the variances under the optimal allocation are equal and we obtain an explicit expression for the optimal allocation by minimizing $\mathrm{Var}(\widehat{\theta}_1)$ setting $p_{01}=p_{11}=1/2$. 
The resulting formula for the allocation proportions in period 2  are:
\begin{eqnarray*}
	p_{22} &=&
	1 + \\ &&\frac{1}{4 \sqrt{3} \sqrt{a (1 - r_1)}}  \left(\sqrt{-b} - \sqrt{
		4 + 8 a + a^2 - 12 (2 + a) r_1 + 
		12 r_1\sqrt{3} \sqrt{\frac{a^3 (1 - r_1)}{-b}}  + 9 r_1^2}\right)
	\\ 
	p_{12} &=&\left(1-\sqrt{4 r_1 p_{22}/(1-r_1)-r_1+4 p_{22}^2/(1-r_1)^2-4 p_{22}/(1-r_1)+1}-r_1\right)/(2(1-r_1)),
\end{eqnarray*}
where $a=\sqrt[3]{9 r_1 (3 (r_1-4) r_1+4)+6 \sqrt{3} \sqrt{r_1 (16-9 r_1 (3 (r_1-4) r_1+8))}+8}$ and $ b=6 (a-4) r_1+(a-2)^2+9 r_1^2$.

\begin{figure}[h!]
	\centering
	\includegraphics[width=0.6\linewidth]{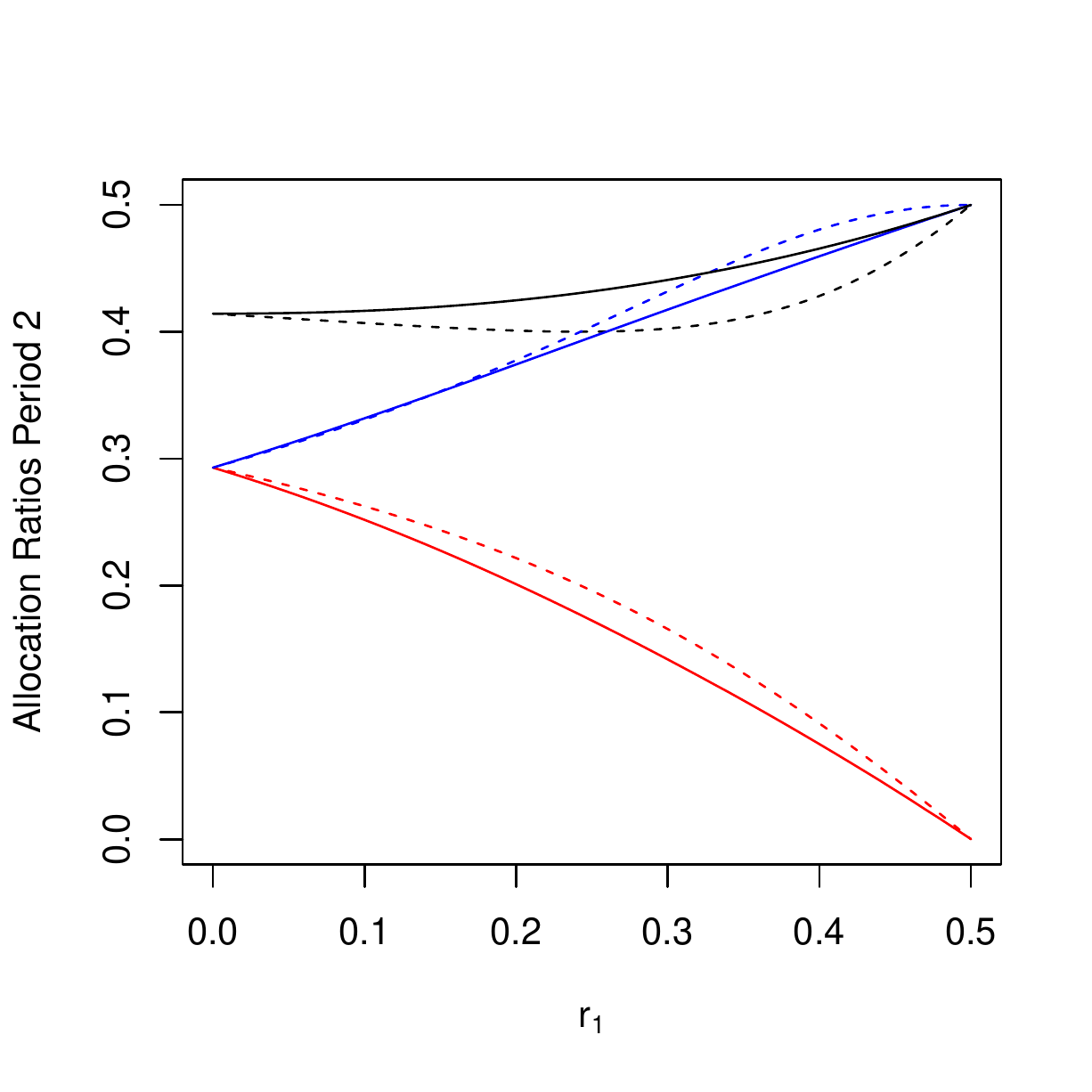}
	\caption{
		Optimal allocation probabilities $r_{i,2}/r_2,i=0,1$ in period 2 as function of $r_1$ for trials with two-periods ($r_3=0$).   Black, red and blue lines represent the allocation rates to control,  arm 1 and  arm 2, respectively. Solid lines refer to analysis with concurrent controls only, and dashed lines to analysis with concurrent and non-concurrent controls.  }
	\label{fig:case2}
\end{figure}

Figure \ref{fig:case2} displays the optimal allocations as a function of $r_1$ both, for trials using non-concurrent controls (dashed lines) and using only concurrent controls (solid lines). Comparing the period 2 allocation rates of the optimal designs with non-concurrent and only concurrent controls, one sees that the latter allocates more patients to the control group. 

\subsection{Case 3: Given sample sizes in periods 1 and 2\label{sec:ncc_case3}}
We now assume that $r_1>0$ and $r_2<1$ are fixed (and not optimised) corresponding to a three-period platform trial in which arm 2 enters later and arm 1 finishes before arm 2 does (Figure \ref{fig:3armspaper}). 
As for concurrent controls, we distinguish 3 situations: (i) If, $r_1> 1/2$, as in Case 2,  the optimal design allocates all patients $j>N_1$ equally to treatment 2 and control, i.e., for $p_{0,1}=p_{1,1}=p_{0,2}=p_{2,2}=p_{0,3}=p_{2,3}=1/2$, 
and $p_{1,2}=0$. (ii) As for concurrent controls, if $r_{1}+r_{2}\leq 1/2$, then for all allocations, the maximum variance is the variance of $\hat \theta_1$ (inclusion of non-concurrent controls can only reduce the variance of $\hat \theta_2$) and therefore the optimal design allocates all observations in period 2 to treatment 1. (iii) If $r_1< 1/2$ and  $r_{1}+r_{2}> 1/2$, we minimise $\max(\mathrm{Var}(\widehat{\theta}_1),\mathrm{Var}(\widetilde\theta_2))$, by  minimizing  $\mathrm{Var}(\widehat{\theta}_1)$ in $\mathbf{p}$ for fixed $r_1$ and $r_2$. In this case, the two variances are equal under the optimal allocation.  
As we could not obtain an analytical solution for the optimization problem in this case, we minimized $\mathrm{Var}(\widehat{\theta}_1)$ in the remaining free variables  $p_{1,2},p_{2,2}$ under the constraint of equal variances with the Mathematica function FindMinimum or the r-package nloptr \cite{nloptr} using the sequential quadratic programming algorithm for nonlinearly constrained gradient-based optimization \cite{kraft1988software}. See Section C in the Supplementary Material. 
Numerical solutions for the optimal allocations in period 2 are shown in dashed lines in Figure \ref{fig:case3}. 
As expected, the larger $r_1$, the larger the difference between designs utilising non-concurrent controls compared to designs with concurrent controls only, as the size of the non-concurrent control group is then larger. The pattern of the optimal allocation rates is similar to the scenario where concurrent controls only are used. However, the ratio of patients assigned to the control group is lower when non-concurrent controls are used. Furthermore, it may be lower than the number of patients assigned to arm 1 or 2 (which does not occur with concurrent controls).



\section{Example and simulation study} \label{sec:casestudy}

We illustrate the optimal allocations in platform trials by means of a  phase  II  placebo-controlled  trial  in primary  hypercholesterolemia \cite{roth2012atorvastatin}. 
In this trial, the goal was to evaluate the efficacy of 80 mg of the antibody atorvastatin with SAR236553 as compared to atorvastatin alone.  Additionally, there was interest in evaluating other doses and combinations, in particular, to investigate the efficacy of 10 mg of atorvastatin plus SAR236553 compared to atorvastatin alone. 
The primary endpoint was the percent change in calculated LDL cholesterol from baseline. 
Patients were randomly assigned according to a 1:1:1 allocation to receive a 80 mg of atorvastatin plus SAR236553, 10 mg of atorvastatin plus SAR236553, or 80 mg of atorvastatin plus placebo.

We revisit the design of this trial to discuss three allocation strategies: equal allocation (1:...:1), square root of $k$ (1:...:$\sqrt{k}$) and the proposed optimal allocations.
We assume the total sample size as in the original trial, $N=92$, and
mean responses and variances  in the control group according as observed in the trial (i.e., a mean of $4.94$ with variance $1$ in the control), and consider means equal to $5.66$ and variance $1$ for the experimental treatment arms, which lead to $80\%$ power at $0.025$ significance level in a multi-arm design setting using the square root of $k$ allocation. 
Furthermore, we considered three designs depending on the entry and end of arm 2 in the trial, that is: 
\begin{enumerate}
	\item Design with one period only (that is, multi-arm design), and thus with sample sizes per period $N_1=N$ and $N_2=N_3=0$ as Case 1 (in Section \ref{sec:case1}).
	\item Design with two periods (arm 2 starts later, but arms 1 and 2 finish at the same time),  as Case 2 (in Section \ref{sec:case2}), assuming a sample size of $N_1=N/4$ in period 1 and $N_2=3N/4$ in period 2.
	\item Design with three periods (arm 2 starts later and finishes after arm 1 does),  as Case 3 (Section \ref{sec:case3}), considering two situations: $N_1=N_2=N_3=N/3$, and $N_1=N/3$, $N_2=4N/9$ and $N_3=2N/9$.
\end{enumerate} 
After rounding, we obtain, e.g., for the three-period design,  $N_1=N_3=31$, and $N_2=30$. We consider the analysis for comparing groups based on concurrent data only. For each design configuration and allocation strategy, we describe the sample size per period and arm. Furthermore, we compare the statistical individual power for each treatment control comparison and also evaluate the variance of the estimates by simulating $100000$ trials. 
In the main manuscript, we report simulation results where no time trends were assumed. In the Supplementary Material (Section B.3), results for trials with time trends are presented.

We start with the standard multi-arm design (Case 1). In this case, as discussed before, the optimal allocation coincides with the root of $k$ rule  (see Table 1 in the Supplementary Material). The power of testing treatments against control is the same in both designs, equal and optimal allocations. When comparing the power of the design with optimal allocations against the design with 1:1 allocation, we can see that there is an increase in power when using the optimal design and the confidence intervals (CIs) are narrower (see Table \ref{sim_power}). However, it could also be noted that in this example, the difference is not substantial. 

Suppose now that the second arm enters while the trial is ongoing, but that it is assumed to end with the first arm. This situation would lead to a two-period design. Tables 2 in the Supplementary Material shows the sample sizes per arm and period. We can observe that the designs using optimal allocations and using  the root of $k$ rule differ. The optimal strategy allocates fewer patients to the first arm in the second period, and more to the second arm in the second period, while maintaining equal allocation between arm 1 and control in the first period as is the case for the one-to-one and square root of $k$ allocations. 
When comparing the powers, the results are as expected considering the optimization strategy: the power of testing the efficacy of arm 1 versus control is larger than the power of the comparison regarding arm 2 when using 1:1 allocation and square root of $k$ allocation (see Table \ref{sim_power}). Under the optimal allocation strategy, the power and standard errors of effect estimates for both treatment control comparisons are the same. As we minimize maximum variance, we see that the maximum power is smaller under the optimal allocation. Similarly,  the (maximum) width of the CIs for the optimal design is smaller than the maximum width of the CI of the other designs. 

Finally, consider a design in which arm 2 enters after arm 1, and in addition, the timing when arm 1 ends is fixed at some point before the end of the trial. For the resulting three-period trial,  we consider two scenarios. In the first, the total sample sizes for periods 1 and 3 are assumed to be equal. Note that this also implies that the total sample sizes for arms 1 and 2 are equal. Then, in the second period, the optimal allocation is the square root of $k$ allocation. When increasing the sample size of period 2 (such that $r_3<r_1$) this is no longer the case. Moreover, as the period where the control arm is shared increases, with the optimal design also the power increases for larger $r_2$ (see Table \ref{sim_power}). Under both choices of $r_2$, the (minimum) power of the optimal design is larger than the minimum power using the 1:1  and square root of $k$ allocation strategies. See Tables 2 and 3  for the sample size per arm and period assumed for each allocation strategy. 

We also evaluated the type 1 error under the different allocation strategies. In all the considered designs, the type 1 error is controlled (see Table 5 in the Supplementary Material), even if there are time trends (see Section B.3 in the Supplementary Material). However, in the latter case, the variances might slightly deviate because the time trends may increase the variability in the treatment groups. 

\begin{table}

	\caption{Power for each design according to the allocation strategy. Here $r_1$ and $r_2$ are the proportion of patients allocated to periods 1 and 2, respectively; ``one'' denotes one-to-one allocation, ``opt'' denotes optimal allocation and ``sqrt''  denotes the square root of $k$ allocation; ``Power A$_i$'' is the estimated power when testing A$_i$ against control ($i=1,2$), and ``CI Width A$_i$'' refers to the width of the confidence interval for the treatment effect of arm A$_i$ compared to  control. Note that the optimal allocation coincides with the square root of $k$ in the 1-period design and 3-period design with $N_1=N_2=N_3=N/3$. }\label{sim_power}
	\centering\small
	\begin{tabular}[t]{l|r|r|l|r|r|r|r}
		\toprule
		Design & $r_1$ & $r_2$ & Allocation & Power  & Power  & CI Width  & CI Width \\
		&  &  &  &  $A1$ &  $A2$ & $A1$ &  $A2$\\
		\toprule
		\multirow{2}{*}{1-period} & 1.000 & 0.000 & one & 0.798 & 0.798 & 1.106 & 1.106  \\
		\cline{2-8}
		& 1.000 & 0.000 & opt (=sqrt) & 0.805 & 0.805 & 1.100 & 1.100  \\
		\toprule
		\multirow{3}{*}{2-period} & 0.250 & 0.750 & one & 0.844 & 0.671 & 1.006 & 1.006  \\
		\cline{2-8}
		& 0.250 & 0.750 & opt & 0.772 & 0.757 & 0.998 & 0.998  \\
		\cline{2-8}
		& 0.250 & 0.750 & sqrt & 0.851 & 0.681 & 0.997 & 0.997  \\
		\toprule
		\multirow{2}{*}{3-period} & 0.337 & 0.326 & one & 0.720 & 0.722 & 1.026 & 1.151  \\
		\cline{2-8}
		& 0.337 & 0.326 & opt (=sqrt) & 0.725 & 0.724 & 1.085 & 1.096 \\
		\toprule
		\multirow{3}{*}{3-period} & 0.337 & 0.446 & one & 0.782 & 0.687 & 0.947 & 1.173 \\
		\cline{2-8}
		& 0.337 & 0.446 & opt & 0.737 & 0.730 & 1.038 & 1.055  \\
		\cline{2-8}
		& 0.337 & 0.446 & sqrt & 0.784 & 0.682 & 0.939 & 1.157 \\
		\bottomrule
	\end{tabular}
\end{table}

\begin{table}[!htb]
	\caption{Sample size distribution per arm and period according to the allocation strategy for a three-period design where arm 2 starts later and finishes after arm 1 does, and with  $N_1=N_2=N_3=N/3$. }
	\small\setlength{\tabcolsep}{1pt}
	\begin{minipage}{.33\linewidth}
		\centering
		
		\label{tab:first_table_3ps}
		\medskip
		\begin{tabular}[t]{l|r|r|r}
			\hline
			& Period 1 & Period 2 & Period 3\\
			\hline
			Arm 2 & 0 & 10 & 16\\
			\hline
			Arm 1 & 16 & 10 & 0\\
			\hline
			Control & 16 & 10 & 16\\
			\hline
		\end{tabular}
		\vspace{5mm}
		\subcaption{One-to-one}
	\end{minipage}\hfill
	\begin{minipage}{.3\linewidth}
		\centering
		
		\label{tab:second_table_3ps}
		\medskip
		\begin{tabular}[t]{l|r|r|r}
			\hline
			& Period 1 & Period 2 & Period 3\\
			\hline
			Arm 2 & 0 & 9 & 16\\
			\hline
			Arm 1 & 16 & 9 & 0\\
			\hline
			Control & 16 & 12 & 16\\
			\hline
		\end{tabular}
		\vspace{5mm}
		\subcaption{$\sqrt{k}$ allocation}
	\end{minipage}\hfill
	\begin{minipage}{.3\linewidth}
		\centering
		
		\label{tab:third_table_3ps}
		\medskip
		\begin{tabular}[t]{l|r|r|r}
			\hline
			& Period 1 & Period 2 & Period 3\\
			\hline
			Arm 2 & 0 & 9 & 16\\
			\hline
			Arm 1 & 16 & 9 & 0\\
			\hline
			Control & 16 & 12 & 16\\
			\hline
		\end{tabular} 
		\vspace{5mm}
		\subcaption{Optimal allocations}
	\end{minipage} 
\end{table}

\begin{table}[!htb]
	\caption{Sample size distribution per arm and period according to the allocation strategy for a three-period design where arm 2 starts later and finishes after arm 1 does, and with  $N_1=N/3$, $N_2=2(N-N_1)/3$ and $N_3=(N-N_1)/3$.}
	\small\setlength{\tabcolsep}{1pt}
	\begin{minipage}{.33\linewidth}
		\centering
		
		\label{tab:first_table_3pns}
		\medskip
		\begin{tabular}[t]{l|r|r|r}
			\hline
			& Period 1 & Period 2 & Period 3\\
			\hline
			Arm 2 & 0 & 14 & 10\\
			\hline
			Arm 1 & 16 & 14 & 0\\
			\hline
			Control & 16 & 14 & 10\\
			\hline
		\end{tabular}
		\vspace{5mm}
		\subcaption{One-to-one}
	\end{minipage}\hfill
	\begin{minipage}{.3\linewidth}
		\centering
		
		\label{tab:second_table_3pns}
		\medskip
		\begin{tabular}[t]{l|r|r|r}
			\hline
			& Period 1 & Period 2 & Period 3\\
			\hline
			Arm 2 & 0 & 12 & 10\\
			\hline
			Arm 1 & 16 & 12 & 0\\
			\hline
			Control & 16 & 17 & 10\\
			\hline
		\end{tabular}
		\vspace{5mm}
		\subcaption{$\sqrt{k}$ allocation}
	\end{minipage}\hfill
	\begin{minipage}{.3\linewidth}
		\centering
		
		\label{tab:third_table_3pns}
		\medskip
		\begin{tabular}[t]{l|r|r|r}
			\hline
			& Period 1 & Period 2 & Period 3\\
			\hline
			Arm 2 & 0 & 16 & 10\\
			\hline
			Arm 1 & 16 & 8 & 0\\
			\hline
			Control & 16 & 17 & 10\\
			\hline
		\end{tabular}
		\vspace{5mm}
		\subcaption{Optimal allocations}
	\end{minipage}  
\end{table}


\section{Discussion}  

We derived optimal allocation rules for platform trials, minimizing the maximum variance of the treatment effect estimators of treatment-control comparisons. The most efficient design is the multi-arm trial, where all treatments start and end at the same time. However, this may not be feasible if, e.g., not all treatments are available at the start. Under the assumption that some treatments enter the trial at a later time point, we showed that in the optimal design, all treatments finish at the end of the platform trial. This again, however, will in general not be practical. Therefore, we considered optimal designs under constraints on the entry and exit times of treatment arms.

In contrast to earlier work, we performed the optimisation assuming that the analyses to compare the efficacy of treatments against control are adjusted with the categorical factor time period. These periods are defined by the time intervals in which no arm enters or leaves the study. Adjusted analyses are recommended in this case to avoid potential biases caused by time trends. Note that the optimal allocation depends on the chosen analysis model. 

In this work, we assumed that the total sample size of the platform trial is known.  Although the total sample size is only a scaling factor in the optimisation problem, this assumption might be unrealistic in this type of trials. However, an equivalent strategy to the one followed to optimise the allocation rates would be to consider a targeted minimum precision (corresponding to the maximum variance) for the estimates of the treatment effects, assuming that these are equal, and minimise the total sample size of the trial. To this end, we need to search for the total sample sizes $N$ such that the minimum precision of the corresponding optimal design is equal to the target precision. As the optimised minimum precision is monotonic in $N$, this corresponds to a simple numerical root finding.


Changing the allocation ratio in a clinical trial has been controversially discussed in the context of response adaptive randomisation \cite{Korn2022,proschan2020resist}. The main concern is the need to adjust for potential time trends, which leads to statistical inefficiency. Also, for platform trials, we saw that the statistically most efficient design is a multi-arm trial, where allocation ratios stay constant over time. However, if it is not feasible to start and complete all arms at the same time, but some arms enter the trial at a later time point, or complete recruitment before the end of the platform trial, a change in the allocation ratios is optimal, also for analysis procedures that adjust for the time trends.  

In this manuscript, we optimised the allocation ratios under the assumption that the times when the second treatment enters or leaves the platform are known. In many realistic settings, these times may, however, be unknown at the start of the trial. In the  resulting, optimal trial designs, however, the optimal allocation ratio in the first period is always 1:1 randomisation and, thus, does not depend on $r_1$ and $r_2$. The optimal allocation ratios for period 2 can be computed when treatment 2 enters the platform and $r_2$ determining when treatment 1 should be completed is fixed. This holds for both, analysis including non-concurrent controls and concurrent controls only. 
If it is desirable to use equal allocation among the experimental treatment arms, one can choose $r_2$ such that the optimal design satisfies this condition (see Figure \ref{fig:case3}). For the analysis using concurrent controls only, the optimal allocation in period 2 is then a $1:1:\sqrt{2}$ allocation.

Our manuscript aims to advance the understanding of optimal allocation principles in platform trials. To accomplish this, we centered our focus on a trial design with two experimental arms and a shared control arm.
The derivation of optimal allocation rules for platform trials with more than two arms is an open problem and subject to further research. 

\section{Supplementary Material}
Web Supplementary Material, including the derivations of the variance of the treatment effect estimators and additional results from the case study, is available with this paper. Mathematica and R code to reproduce the results are available at \url{https://github.com/MartaBofillRoig/Allocation}.

\section*{Acknowledgments}

The authors are members the EU Patient-centric clinical trial platform (EU-PEARL). EU-PEARL (EU Patient-cEntric clinicAl tRial pLatforms) project has received funding from the Innovative Medicines Initiative (IMI) 2 Joint Undertaking (JU) under grant agreement No 853966. This Joint Undertaking receives support from the European Union’s Horizon 2020 research and innovation programme and EFPIA and Children's Tumor Foundation, Global Alliance for TB Drug Development non-profit organisation, Springworks Therapeutics Inc. This publication reflects the authors' views. Neither IMI nor the European Union, EFPIA, or any Associated Partners are responsible for any use that may be made of the information contained herein.



\bibliography{refs}

\newpage

...



\section*{Supplementary material}

# Supplementary material of "Optimal allocation strategies in platform trials"

Marta Bofill Roig[1], Ekkehard Glimm[2], Tobias Mielke[3], and Martin Posch[*1]

[1]Section for Medical Statistics, Center for Medical Data Science, Medical University of Vienna, Vienna
[2]Advanced Methodology and Data Science, Novartis Pharma AG, Basel
[3]Statistics and Decision Sciences, Janssen-Cilag GmbH

# Contents

---

*martin.posch@meduniwien.ac.at

In this supplementary material, we provide the calculations of the variance estimators (Section A) and additional results concerning the simulation study (Section B). Mathematica and R code to reproduce the results are available at `https://github.com/MartaBofillRoig/Allocation`. However, for the reader's convenience, we also include the Mathematica files at the end of this document.

# A   Variance derivations

Consider the same notation used in Section 2 of the article and let $n_{i,s}, i = 0, 1, 2, s = 1, 2, 3$ be the sample sizes in arm $i$, period $s$ of the platform trial design.

## A.1   Variance for effect estimators using concurrent controls only

Consider the period-wise treatment effect estimators of the form

$$\widehat{\theta}_i = \sum_{s=i,i+1} w_{i,s} \cdot \hat{\theta}_{i,s}$$

where $\hat{\theta}_{i,s} = \bar{y}_{i,s} - \bar{y}_{0,s}$ is the treatment effect estimate in period $s$ for arm $i$ ($i = 1, 2$), where $\bar{y}_{i,s}$ and $\bar{y}_{0,s}$ are the sample means in period $s$ for experimental and control arms, and the weights are given by

$$w_{1,s} = \frac{\frac{1}{\sigma_{1,s}^2}}{\frac{1}{\sigma_{1,1}^2} + \frac{1}{\sigma_{1,2}^2}} \quad \text{and} \quad w_{2,s} = \frac{\frac{1}{\sigma_{2,s}^2}}{\frac{1}{\sigma_{2,2}^2} + \frac{1}{\sigma_{2,3}^2}}, \tag{1.a}$$

where $\sigma_{i,s}^2 = \text{Var}(\hat{\theta}_{i,s})$ denotes the variances of the estimates per period, and is given by $\sigma_{i,s}^2 = \sigma^2(1/n_{i,s} + 1/n_{0,s})$.

Then, the variance of the estimator is then given by:

$$
\begin{aligned}
\text{Var}(\widehat{\theta}_i) &= \sum_{s=i,i+1} w_{i,s}^2 \sigma_{i,s}^2 = \sum_{s=i,i+1} \left( \frac{\frac{1}{\sigma_{i,s}^2}}{\frac{1}{\sigma_{i,i}^2} + \frac{1}{\sigma_{i,i+1}^2}} \right)^2 \sigma_{i,s}^2 = \sum_{s=i,i+1} \frac{\frac{1}{\sigma_{i,s}^2}}{\left( \frac{1}{\sigma_{i,i}^2} + \frac{1}{\sigma_{i,i+1}^2} \right)^2} = \frac{1}{\frac{1}{\sigma_{i,i}^2} + \frac{1}{\sigma_{i,i+1}^2}} \\
&= \sigma^2 \cdot \left( \frac{1}{\frac{1}{n_{i,i}} + \frac{1}{n_{0,i}}} + \frac{1}{\frac{1}{n_{i,i+1}} + \frac{1}{n_{0,i+1}}} \right)^{-1} = \sigma^2 \cdot \left( \frac{n_{i,i} n_{0,i}}{n_{i,i} + n_{0,i}} + \frac{n_{i,i+1} n_{0,i+1}}{n_{i,i+1} + n_{0,i+1}} \right)^{-1} \\
&= \frac{\sigma^2}{N} \cdot \left( r_i \frac{p_{i,i} p_{0,i}}{p_{i,i} + p_{0,i}} + r_{i+1} \frac{p_{i,i+1} p_{0,i+1}}{p_{i,i+1} + p_{0,i+1}} \right)^{-1}.
\end{aligned}
$$

as $p_{i,s} = \frac{n_{i,s}}{r_s N_s}$ and $r_s = \frac{N_s}{N}$.

## A.2   Variance for the effect estimator using concurrent and non-concurrent controls

In this section, we first derive the expression of the effect estimators using concurrent and non-concurrent controls and then give the variance expressions. Note that in what follows, the treatment effect estimator for arm 1 uses concurrent and non-concurrent controls as well, as arm 2 does. However, in the case of arm 1, non-concurrent controls are those patients allocated to the control in period 3. For the purpose of optimal allocations in this paper, we are only interested in the estimator of arm 2, but we give here the expressions of the estimator of arm 1 for completeness.

In Section A.2.1, we derive the expressions of the estimators and the corresponding variances when utilising non-concurrent controls together with concurrent controls. In Sections A.2.2 and A.2.3, we give the expressions for the particular case of a platform trial with two-periods and three-periods, respectively. The expressions given in the later sections are the ones we use in Section 4 of the manuscript.

### A.2.1 Effect estimator using concurrent and non-concurrent controls

We assume a linear model $E(\mathbf{y}) = \boldsymbol{X}\boldsymbol{\theta}$ where $\boldsymbol{\theta} = (\mu_1, \mu_2, \mu_3, \theta_1, \theta_2)'$ is the parameter vector with baseline stagewise responses $\mu_s$ and treatment effects $\theta_i$. $\boldsymbol{X}$ is the design matrix and $\mathbf{y}$ the vector of independent observations with a constant variance $\sigma^2$ (without loss of generality, $\sigma = 1$).

The Gauss-Markov (best linear unbiased) estimator of $\boldsymbol{\theta}$ is $\hat{\boldsymbol{\theta}} = (\boldsymbol{X}'\boldsymbol{X})^{-1}\boldsymbol{X}'\mathbf{y}$. It is best linear unbiased in the sense that for any other estimate $\tilde{\boldsymbol{\theta}} = \boldsymbol{A}\boldsymbol{y}$ with $E(\tilde{\boldsymbol{\theta}}) = \boldsymbol{\theta}$, $Cov(\tilde{\boldsymbol{\theta}}) - Cov(\hat{\boldsymbol{\theta}})$ is positive semidefinite which in turn implies that the variance of any component of $\hat{\boldsymbol{\theta}}$ is at a minimum among these unbiased linear estimates. We have $Cov(\hat{\boldsymbol{\theta}}) = \sigma^2 \left(\boldsymbol{X}'\boldsymbol{X}\right)^{-1}$.

In the specific case here, the design matrix is

$$\boldsymbol{X} = \begin{pmatrix} \mathbf{1}_{n_{0,1}} & \mathbf{0} & \mathbf{0} & \mathbf{0} & \mathbf{0} \\ \mathbf{1}_{n_{1,1}} & \mathbf{0} & \mathbf{0} & \mathbf{1}_{n_{1,1}} & \mathbf{0} \\ \mathbf{0} & \mathbf{1}_{n_{0,2}} & \mathbf{0} & \mathbf{0} & \mathbf{0} \\ \mathbf{0} & \mathbf{1}_{n_{1,2}} & \mathbf{0} & \mathbf{1}_{n_{1,2}} & \mathbf{0} \\ \mathbf{0} & \mathbf{1}_{n_{2,2}} & \mathbf{0} & \mathbf{0} & \mathbf{1}_{n_{2,2}} \\ \mathbf{0} & \mathbf{0} & \mathbf{1}_{n_{0,3}} & \mathbf{0} & \mathbf{0} \\ \mathbf{0} & \mathbf{0} & \mathbf{1}_{n_{2,3}} & \mathbf{0} & \mathbf{1}_{n_{2,3}} \end{pmatrix}$$

where $\mathbf{1}_x$ denotes a vector of $x$ ones. Hence,

$$\boldsymbol{X}'\boldsymbol{X} = \begin{pmatrix} N_1 & 0 & 0 & n_{1,1} & 0 \\ 0 & N_2 & 0 & n_{1,2} & n_{2,2} \\ 0 & 0 & N_3 & 0 & n_{2,3} \\ n_{11} & n_{1,2} & 0 & n_{1,1} + n_{1,2} & 0 \\ 0 & n_{2,2} & n_{2,3} & 0 & n_{2,2} + n_{2,3} \end{pmatrix}$$

where $N_s$ is the sample size per period, that is, $N_1 = n_{0,1} + n_{1,1}$, $N_2 = n_{0,2} + n_{1,2} + n_{2,2}$, $N_3 = n_{0,3} + n_{2,3}$. We can use the technique of inverting block-partitioned matrices here. If we write

$$\boldsymbol{A} = \begin{pmatrix} N_1 & 0 & 0 \\ 0 & N_2 & 0 \\ 0 & 0 & N_3 \end{pmatrix},$$

$$\boldsymbol{C} = \begin{pmatrix} n_{1,1} + n_{1,2} & 0 \\ 0 & n_{2,2} + n_{2,3} \end{pmatrix},$$

and

$$\boldsymbol{B} = \begin{pmatrix} n_{1,1} & 0 \\ n_{1,2} & n_{2,2} \\ 0 & n_{2,3} \end{pmatrix},$$

then, by this technique, we obtain

$$cov(\widetilde{\theta}_1, \widetilde{\theta}_2) = \left(\boldsymbol{C} - \boldsymbol{B}'\boldsymbol{A}^{-1}\boldsymbol{B}\right)^{-1} = f^{-1} \cdot \begin{pmatrix} n_{2,2} + n_{2,3} - \frac{n_{2,2}^2}{N_2} - \frac{n_{2,3}^2}{N_3} & \frac{n_{1,2}n_{2,2}}{N_2} \\ \frac{n_{1,2}n_{2,2}}{N_2} & n_{1,1} + n_{1,2} - \frac{n_{1,1}^2}{N_1} - \frac{n_{1,2}^2}{N_2} \end{pmatrix} \quad (1.b)$$

where $f = \left(n_{1,1} + n_{1,2} - \frac{n_{1,1}^2}{N_1} - \frac{n_{1,2}^2}{N_2}\right) \cdot \left(n_{2,2} + n_{2,3} - \frac{n_{2,2}^2}{N_2} - \frac{n_{2,3}^2}{N_3}\right) - \left(\frac{n_{1,2}n_{2,2}}{N_2}\right)^2$.

Regarding the point estimates, it is possible to simplify to

$$\begin{pmatrix} \widetilde{\theta}_1 \\ \widetilde{\theta}_2 \end{pmatrix} = \left(\boldsymbol{C} - \boldsymbol{B}'\boldsymbol{A}^{-1}\boldsymbol{B}\right)^{-1} \left(-\boldsymbol{B}'\boldsymbol{A}^{-1} \vdots \boldsymbol{I}_2\right) \begin{pmatrix} N_1\bar{y}_{.,1} \\ N_2\bar{y}_{.,2} \\ N_3\bar{y}_{.,3} \\ n_{1,1}\bar{y}_{1,1} + n_{1,2}\bar{y}_{1,2} \\ n_{2,2}\bar{y}_{2,2} + n_{1,2}\bar{y}_{2,3} \end{pmatrix} = \quad (1.c)$$
$$\left(\boldsymbol{C} - \boldsymbol{B}'\boldsymbol{A}^{-1}\boldsymbol{B}\right)^{-1} \begin{pmatrix} n_{1,1}(\bar{y}_{1,1} - \bar{y}_{.,1}) + n_{1,2}(\bar{y}_{1,2} - \bar{y}_{.,2}) \\ n_{2,2}(\bar{y}_{2,2} - \bar{y}_{.,2}) + n_{2,3}(\bar{y}_{2,3} - \bar{y}_{.,3}) \end{pmatrix}$$

where $\bar{y}_{i,s}$ is the pooled mean per arm and period, and $\bar{y}_{.,s}$ is the pooled mean per period. Note, however, that $\widetilde{\theta}_i$ cannot, in general, be represented as a weighted average of stagewise treatment effect contrasts, i.e. as $\sum_{i=1}^{2} \sum_{s=1}^{3} w_{i,s} \cdot (\bar{y}_{i,s} - \bar{y}_{0,s})$.

We focus on the treatment effect estimator for treatment 2 only, given that this is the arm in which we use non-concurrent controls. Assume now that $N_s = r_s N$ is the sample size in period $s$ and $r_{i,s} = \frac{n_{i,s}}{N}$ is the fraction of patients in arm $i = 0, 1, 2$ and period $s$ among all patients such that $r_s = \sum_{i \in I_s} r_{i,s}$ and $r_1 + r_2 + r_3 = 1$. $\mathrm{Var}(\widetilde{\theta}_1)$ and $\mathrm{Var}(\widetilde{\theta}_2)$ are the diagonal of (1.b). With this notation, we can write the variance as

$$\mathrm{Var}(\widetilde{\theta}_2) = \frac{r_1 p_{1,1} \left(1 - p_{1,1}\right) + r_2 p_{1,2} \left(1 - p_{1,2}\right)}{\left(r_1 p_{1,1} \left(1 - p_{1,1}\right) + r_2 p_{1,2} \left(1 - p_{1,2}\right)\right) \left(r_3 p_{2,3} \left(1 - p_{2,3}\right) + r_2 p_{2,2} \left(1 - p_{2,2}\right)\right) - r_2^2 p_{1,2}^2 p_{2,2}^2} \cdot \frac{1}{N} \quad (1.\mathrm{d})$$

where $p_{i,s} = \frac{r_{i,s}}{r_s}$ is the proportion of patients in treatment group $i$ of period $s$.

We assume that $r_1$ is fixed. Minimizing $\mathrm{Var}(\widetilde{\theta}_i)$ is equivalent to maximizing $\mathrm{Var}(\widetilde{\theta}_i)^{-1}$. Denoting by $q_{js} = p_{js}(1 - p_{js})$, we obtain

$$\mathrm{Var}(\widetilde{\theta}_2)^{-1} \propto r_3 q_{23} + r_2 q_{22} - \frac{r_2^2 p_{12}^2 p_{22}^2}{r_1 q_{11} + r_2 q_{12}}. \quad (1.\mathrm{e})$$

### A.2.2 Two-period trial

Consider a two-period trial, where arm 2 enters later and finishes at the same time that arm 1, as illustrated in Figure 3 of the paper. The treatment effect estimator for treatment 2 using non-concurrent controls can then be written as

$$\widetilde{\theta}_2 = \hat{\theta}_{2,2} + \rho(\hat{\theta}_{1,1} - \hat{\theta}_{1,2})$$

where

$$\rho = \frac{\frac{1}{n_{0,2}}}{\frac{1}{n_{0,1}} + \frac{1}{n_{0,2}} + \frac{1}{n_{1,1}} + \frac{1}{n_{1,2}}}$$

The variance of the estimator $\widetilde{\theta}_2$ is then

$$
\begin{aligned}
\mathrm{Var}(\widetilde{\theta}_2) &= \mathrm{Var}(\hat{\theta}_{2,2}) + \rho^2 \mathrm{Var}(\hat{\theta}_{1,1}) + \rho^2 \mathrm{Var}(\hat{\theta}_{1,2}) - 2\rho \mathrm{Cov}(\hat{\theta}_{2,2}, \hat{\theta}_{1,2}) = \sigma_{2,2}^2 + \rho^2(\sigma_{1,1}^2 + \sigma_{1,2}^2) - 2\rho \frac{\sigma^2}{n_{0,2}} \\
&= \sigma^2 \left( \frac{1}{n_{2,2}} + \frac{1}{n_{0,2}} \right) - 2\rho\sigma^2 \frac{1}{n_{0,2}} + \rho^2 \sigma^2 \left( \frac{1}{n_{1,1}} + \frac{1}{n_{0,1}} + \frac{1}{n_{1,2}} + \frac{1}{n_{0,2}} \right) \\
&= \sigma^2 \left( \frac{1}{n_{2,2}} + \frac{1}{n_{0,2}} \right) - 2\sigma^2 \cdot \frac{\left( \frac{1}{n_{0,2}} \right)^2}{\frac{1}{n_{0,1}} + \frac{1}{n_{0,2}} + \frac{1}{n_{1,1}} + \frac{1}{n_{1,2}}} + \sigma^2 \cdot \frac{\left( \frac{1}{n_{0,2}} \right)^2}{\frac{1}{n_{0,1}} + \frac{1}{n_{0,2}} + \frac{1}{n_{1,1}} + \frac{1}{n_{1,2}}} \\
&= \sigma^2 \left( \frac{1}{n_{2,2}} + \frac{1}{n_{0,2}} - \frac{\left( \frac{1}{n_{0,2}} \right)^2}{\frac{1}{n_{0,1}} + \frac{1}{n_{0,2}} + \frac{1}{n_{1,1}} + \frac{1}{n_{1,2}}} \right)
\end{aligned}
$$

By taking into account $p_{i,s} = \frac{n_{i,s}}{r_s N_s}$, then:

$$\mathrm{Var}(\widetilde{\theta}_2) = \frac{\sigma^2}{N} \left( r_2 q_{2,2} - \frac{r_2^2 p_{1,2}^2 p_{2,2}^2}{r_1 q_{1,1} + r_2 q_{1,2}} \right)^{-1} \quad (1.\mathrm{f})$$

### A.2.3 Three-period trial

Suppose now a three-period platform trial in which arm 2 enters later and arm 1 finishes before arm 2 does (see Figure 1 in the paper for an illustration). The treatment effect estimator for comparing arm 2 against control when using the model-based approaches is

$$\widetilde{\theta}_2 \;=\; \omega_{1,1}(\hat{y}_{1,1} - \hat{y}_{.,1}) + \omega_{1,2}(\hat{y}_{1,2} - \hat{y}_{.,2}) + \omega_{2,2}(\hat{y}_{2,2} - \hat{y}_{.,2}) + \omega_{2,3}(\hat{y}_{2,3} - \hat{y}_{.,3}) \tag{1.g}$$

where $\hat{y}_{.,s}$ is the pooled mean per period, and where $\omega_{i,s}$ are given by:

$$
\begin{aligned}
\omega_{1,1} \;=\; & \Big(n_{1,1}(n_{0,1} + n_{1,1})n_{1,2}n_{2,2}(n_{0,3} + n_{2,3})\Big) \cdot \Big(n_{1,1}n_{1,2}(n_{0,2}n_{0,3}n_{2,2} + n_{0,3}n_{2,2}n_{2,3} + n_{0,2}(n_{0,3} + n_{2,2})n_{2,3}) \\
& + n_{0,1}(n_{0,3}n_{1,1}n_{1,2}n_{2,2} + (n_{0,3}n_{1,1}n_{1,2} + n_{1,1}n_{1,2}n_{2,2} + n_{0,3}(n_{1,1} + n_{1,2})n_{2,2})n_{2,3} \\
& + n_{0,2}(n_{1,1} + n_{1,2})(n_{2,2}n_{2,3} + n_{0,3}(n_{2,2} + n_{2,3})))\Big)^{-1}
\end{aligned}
$$

$$
\begin{aligned}
\omega_{1,2} \;=\; & \Big((n_{0,1} + n_{1,1})n_{1,2}^2 n_{2,2}(n_{0,3} + n_{2,3})\Big) \cdot \Big(n_{1,1}n_{1,2}(n_{0,2}n_{0,3}n_{2,2} + n_{0,3}n_{2,2}n_{2,3} + n_{0,2}(n_{0,3} + n_{2,2})n_{2,3}) + \\
& + n_{0,1}(n_{0,3}n_{1,1}n_{1,2}n_{2,2} + (n_{0,3}n_{1,1}n_{1,2} + n_{1,1}n_{1,2}n_{2,2} + n_{0,3}(n_{1,1} + n_{1,2})n_{2,2})n_{2,3} \\
& + n_{0,2}(n_{1,1} + n_{1,2})(n_{2,2}n_{2,3} + n_{0,3}(n_{2,2} + n_{2,3})))\Big)^{-1}
\end{aligned}
$$

$$
\begin{aligned}
\omega_{2,2} \;=\; & \Big(n_{2,2}(n_{1,1}n_{1,2}(n_{0,2} + n_{2,2}) + n_{0,1}(n_{1,1}n_{1,2} + n_{0,2}(n_{1,1} + n_{1,2}) + (n_{1,1} + n_{1,2})n_{2,2}))(n_{0,3} + n_{2,3})\Big) \cdot \\
& \cdot \Big(n_{1,1}n_{1,2}(n_{0,2}n_{0,3}n_{2,2} + n_{0,3}n_{2,2}n_{2,3} + n_{0,2}(n_{0,3} + n_{2,2})n_{2,3}) + \\
& + n_{0,1}(n_{0,3}n_{1,1}n_{1,2}n_{2,2} + (n_{0,3}n_{1,1}n_{1,2} + n_{1,1}n_{1,2}n_{2,2} + n_{0,3}(n_{1,1} + n_{1,2})n_{2,2})n_{2,3} + \\
& + n_{0,2}(n_{1,1} + n_{1,2})(n_{2,2}n_{2,3} + n_{0,3}(n_{2,2} + n_{2,3})))\Big)^{-1}
\end{aligned}
$$

$$
\begin{aligned}
\omega_{2,3} \;=\; & \Big((n_{1,1}n_{1,2}(n_{0,2} + n_{2,2}) + n_{0,1}(n_{1,1}n_{1,2} + n_{0,2}(n_{1,1} + n_{1,2}) + (n_{1,1} + n_{1,2})n_{2,2}))n_{2,3}(n_{0,3} + n_{2,3})\Big) \cdot \\
& \cdot \Big(n_{1,1}n_{1,2}(n_{0,2}n_{0,3}n_{2,2} + n_{0,3}n_{2,2}n_{2,3} + n_{0,2}(n_{0,3} + n_{2,2})n_{2,3}) + \\
& + n_{0,1}(n_{0,3}n_{1,1}n_{1,2}n_{2,2} + (n_{0,3}n_{1,1}n_{1,2} + n_{1,1}n_{1,2}n_{2,2} + n_{0,3}(n_{1,1} + n_{1,2})n_{2,2})n_{2,3} + \\
& + n_{0,2}(n_{1,1} + n_{1,2})(n_{2,2}n_{2,3} + n_{0,3}(n_{2,2} + n_{2,3})))\Big)^{-1}
\end{aligned}
$$

The variance of the estimator $\widetilde{\theta}_2$ is given in (1.d).

# B  Additional simulation results

## B.1  Sample sizes per period and arm

In what follows, we describe the sample size distribution per period and arm according to each design configuration and allocation strategy.

The considered designs are:

1. Design with one period only (that is, multi-arm design), and thus with sample sizes per period $N_1 = N$ and $N_2 = N_3 = 0$.

2. Design with two periods (arm 2 starts later, but arms 1 and 2 finish at the same time), assuming a sample size of $N_1 = N/4$ in period 1 and $N_2 = 3N/4$ in period 2.

3. Design with three periods (arm 2 starts later and finishes after arm 1 does), where considering two situations: $N_1 = N_2 = N_3 = N/3$, and $N_1 = N/3$ $N_2 = 2(N - N_1)/3$ and $N_3 = (N - N_1)/3$.

where $N = 92$. The following tables depict the sample size per arm and period according to each design and with respect to the allocation strategy.

Table 1: Sample size distribution per arm and period according to the allocation strategy for a multi-arm trial design (one-period design).

| | Period 1 | Period 2 | Period 3 |
|---|---|---|---|
| Arm 2 | 31 | 0 | 0 |
| Arm 1 | 31 | 0 | 0 |
| Control | 31 | 0 | 0 |

(a) One-to-one

| | Period 1 | Period 2 | Period 3 |
|---|---|---|---|
| Arm 2 | 27 | 0 | 0 |
| Arm 1 | 27 | 0 | 0 |
| Control | 38 | 0 | 0 |

(b) $\sqrt{k}$ allocation

| | Period 1 | Period 2 | Period 3 |
|---|---|---|---|
| Arm 2 | 27 | 0 | 0 |
| Arm 1 | 27 | 0 | 0 |
| Control | 38 | 0 | 0 |

(c) Optimal allocations

Table 2: Sample size distribution per arm and period according to the allocation strategy for a two-period design where arm 2 starts later and finishes when arm 1 does, and with sample size of $N_1 = N/4$ in period 1 and $N_2 = 3N/4$ in period 2.

| | Period 1 | Period 2 | Period 3 |
|---|---|---|---|
| Arm 2 | 0 | 23 | 0 |
| Arm 1 | 12 | 23 | 0 |
| Control | 12 | 23 | 0 |

(a) One-to-one

| | Period 1 | Period 2 | Period 3 |
|---|---|---|---|
| Arm 2 | 0 | 20 | 0 |
| Arm 1 | 12 | 20 | 0 |
| Control | 12 | 29 | 0 |

(b) $\sqrt{k}$ allocation

| | Period 1 | Period 2 | Period 3 |
|---|---|---|---|
| Arm 2 | 0 | 27 | 0 |
| Arm 1 | 12 | 12 | 0 |
| Control | 12 | 30 | 0 |

(c) Optimal allocations

Table 3: Sample size distribution per arm and period according to the allocation strategy for a three-period design where arm 2 starts later and finishes after arm 1 does, and with $N_1 = N_2 = N_3 = N/3$.

| | Period 1 | Period 2 | Period 3 |
|---|---|---|---|
| Arm 2 | 0 | 10 | 16 |
| Arm 1 | 16 | 10 | 0 |
| Control | 16 | 10 | 16 |

(a) One-to-one

| | Period 1 | Period 2 | Period 3 |
|---|---|---|---|
| Arm 2 | 0 | 9 | 16 |
| Arm 1 | 16 | 9 | 0 |
| Control | 16 | 12 | 16 |

(b) $\sqrt{k}$ allocation

| | Period 1 | Period 2 | Period 3 |
|---|---|---|---|
| Arm 2 | 0 | 9 | 16 |
| Arm 1 | 16 | 9 | 0 |
| Control | 16 | 12 | 16 |

(c) Optimal allocations

Table 4: Sample size distribution per arm and period according to the allocation strategy for a three-period design where arm 2 starts later and finishes after arm 1 does, and with $N_1 = N/3$, $N_2 = 2(N - N_1)/3$ and $N_3 = (N - N_1)/3$.

| | Period 1 | Period 2 | Period 3 |
|---|---|---|---|
| Arm 2 | 0 | 14 | 10 |
| Arm 1 | 16 | 14 | 0 |
| Control | 16 | 14 | 10 |

(a) One-to-one

| | Period 1 | Period 2 | Period 3 |
|---|---|---|---|
| Arm 2 | 0 | 12 | 10 |
| Arm 1 | 16 | 12 | 0 |
| Control | 16 | 17 | 10 |

(b) $\sqrt{k}$ allocation

| | Period 1 | Period 2 | Period 3 |
|---|---|---|---|
| Arm 2 | 0 | 16 | 10 |
| Arm 1 | 16 | 8 | 0 |
| Control | 16 | 17 | 10 |

(c) Optimal allocations

## B.2 Results under the null hypothesis

In this section, we report the results of the simulations under the null hypothesis. In this case, we considered means for the treatment arms equal to 4.94 as in the control group, and variances equal to 1. Table 5 summarizes the results. We can observe that the type 1 error is controlled in all considered scenarios.

## B.3 Results for trials with time trends

In this section, we present the results of the case study for trials with time trends. We considered the designs presented in the previous section, and the same parameter settings as in the paper: a total sample size of $N = 92$, the mean in the control is 4.94 with variance 1, and means equal to 5.66 and variance 1 in the treatment arms. In this simulation, however, we considered step-wise time trends. Specifically, the data are

Table 5: Type 1 error (T1E) rates for each design according to the allocation strategy. Here $r_1$ and $r_2$ are the proportion of patients allocated to periods 1 and 2, respectively; "one" denotes one-to-one allocation, "opt" denotes optimal allocation and "sqrt" denotes the square root of $k$ allocation; "T1E A$_i$" is the estimated type 1 error when testing A$_i$ against control ($i = 1, 2$), and "CI Width A$_i$" refers to the width of the confidence interval for the treatment effect of arm A$_i$ against control.

| Design | $r_1$ | $r_2$ | Allocation | T1E A1 | T1E A2 | CI Width A1 | CI Width A2 |
|--------|-------|-------|------------|--------|--------|-------------|-------------|
| 1-period | 1.000 | 0.000 | one | 0.025 | 0.026 | 1.106 | 1.106 |
|          | 1.000 | 0.000 | opt | 0.025 | 0.026 | 1.099 | 1.100 |
|          | 1.000 | 0.000 | sqrt | 0.025 | 0.025 | 1.100 | 1.100 |
| 2-period | 0.250 | 0.750 | one | 0.025 | 0.025 | 1.006 | 1.006 |
|          | 0.250 | 0.750 | opt | 0.026 | 0.025 | 0.997 | 0.997 |
|          | 0.250 | 0.750 | sqrt | 0.025 | 0.025 | 0.997 | 0.997 |
| 3-period | 0.337 | 0.326 | one | 0.025 | 0.025 | 1.026 | 1.151 |
|          | 0.337 | 0.326 | opt | 0.025 | 0.025 | 1.084 | 1.095 |
|          | 0.337 | 0.326 | sqrt | 0.025 | 0.024 | 1.025 | 1.149 |
| 3-period | 0.337 | 0.446 | one | 0.025 | 0.025 | 0.947 | 1.173 |
|          | 0.337 | 0.446 | opt | 0.025 | 0.026 | 1.038 | 1.055 |
|          | 0.337 | 0.446 | sqrt | 0.025 | 0.026 | 0.938 | 1.156 |

generated according to the model

$$E(Y_j) = \eta_0 + \sum_{i=1,2} \theta_i \cdot I(i_j = i) + f(t_j), \tag{2.h}$$

where $\eta_0$ is the control response in period 1, $\theta_k$ is the treatment effect for arm $k$, $f(\cdot)$ represents the time trend function and $t_j$ is the calendar time when patient $j$ is enrolled in the trial, where $f(j) = \lambda I(j > N_1)$. Here $\lambda$ represents the strength of the time trend, which is considered to be $\lambda = 0.25$ in the simulations.

In the following tables, we present the resulting power and type 1 error for each design and allocation strategy. We can see that the results also hold for trials with time trends and that the type 1 error is maintained.

## C  Mathematica files

The following Mathematica notebook files are now displayed:

- suppmat-CC.nb: Computation on designs with concurrent controls only.

- suppmat-NCC-case2.nb: Computations on case 2 using non-concurrent controls.

- suppmat-NCC-case3.nb: Computations on case 3 using non-concurrent controls.

Mathematica and R code files are also available at `https://github.com/MartaBofillRoig/Allocation`.

Table 6: Power for each design according to the allocation strategy for a trial with stepwise time trends (see (2.h) for the data generation). Here $r_1$ and $r_2$ are the proportion of patients allocated to periods 1 and 2, respectively; "one" denotes one-to-one allocation, "opt" denotes optimal allocation and "sqrt" denotes the square root of $k$ allocation; "Power $A_i$" is the estimated type 1 error when testing $A_i$ against control ($i = 1, 2$), and "CI Width $A_i$" and "Variance $A_i$" refers to the width of the confidence interval and variance, respectively, for the treatment effect of arm $A_i$ against control.

| Design | $r_1$ | $r_2$ | Allocation | Power $A1$ | Power $A2$ | CI Width $A1$ | CI Width $A2$ |
|---|---|---|---|---|---|---|---|
| 1-period | 1.000 | 0.000 | one | 0.799 | 0.796 | 1.105 | 1.106 |
| | 1.000 | 0.000 | opt | 0.805 | 0.805 | 1.099 | 1.099 |
| | 1.000 | 0.000 | sqrt | 0.805 | 0.804 | 1.100 | 1.099 |
| 2-period | 0.250 | 0.750 | one | 0.844 | 0.666 | 1.006 | 1.006 |
| | 0.250 | 0.750 | opt | 0.772 | 0.759 | 0.997 | 0.997 |
| | 0.250 | 0.750 | sqrt | 0.852 | 0.683 | 0.997 | 0.997 |
| 3-period | 0.337 | 0.326 | one | 0.717 | 0.721 | 1.027 | 1.151 |
| | 0.337 | 0.326 | opt | 0.724 | 0.724 | 1.085 | 1.096 |
| | 0.337 | 0.326 | sqrt | 0.726 | 0.725 | 1.025 | 1.150 |
| 3-period | 0.337 | 0.446 | one | 0.782 | 0.687 | 0.947 | 1.173 |
| | 0.337 | 0.446 | opt | 0.737 | 0.727 | 1.038 | 1.055 |
| | 0.337 | 0.446 | sqrt | 0.786 | 0.688 | 0.938 | 1.157 |

Table 7: Type 1 error (T1E) rates for each design according to the allocation strategy for a trial with stepwise time trends (see (2.h) for the data generation). Here $r_1$ and $r_2$ are the proportion of patients allocated to periods 1 and 2, respectively; "one" denotes one-to-one allocation, "opt" denotes optimal allocation and "sqrt" denotes the square root of $k$ allocation; "T1E $A_i$" is the estimated type 1 error when testing $A_i$ against control ($i = 1, 2$); "CI Width $A_i$" and "Variance $A_i$" refers to the width of the confidence interval and variance, respectively, for the treatment effect of arm $A_i$ against control.

| Design | $r_1$ | $r_2$ | Allocation | T1E $A1$ | T1E $A2$ | CI Width $A1$ | CI Width $A2$ |
|---|---|---|---|---|---|---|---|
| 1-period | 1.000 | 0.000 | one | 0.025 | 0.024 | 1.105 | 1.105 |
| | 1.000 | 0.000 | opt | 0.025 | 0.025 | 1.100 | 1.100 |
| | 1.000 | 0.000 | sqrt | 0.025 | 0.025 | 1.099 | 1.100 |
| 2-period | 0.250 | 0.750 | one | 0.026 | 0.024 | 1.006 | 1.006 |
| | 0.250 | 0.750 | opt | 0.025 | 0.025 | 0.997 | 0.997 |
| | 0.250 | 0.750 | sqrt | 0.025 | 0.026 | 0.998 | 0.998 |
| 3-period | 0.337 | 0.326 | one | 0.024 | 0.024 | 1.027 | 1.151 |
| | 0.337 | 0.326 | opt | 0.025 | 0.025 | 1.084 | 1.095 |
| | 0.337 | 0.326 | sqrt | 0.025 | 0.024 | 1.026 | 1.150 |
| 3-period | 0.337 | 0.446 | one | 0.025 | 0.025 | 0.947 | 1.173 |
| | 0.337 | 0.446 | opt | 0.025 | 0.025 | 1.038 | 1.055 |
| | 0.337 | 0.446 | sqrt | 0.025 | 0.025 | 0.939 | 1.157 |

# Supplementary Material of "Optimal allocation strategies in platform trials"

## Design with concurrent controls only

In this file, we provide the derivation of the optimal allocation for trials utilising concurrent controls only. In what follows, we focus on determining the optimal solutions for the Case 3 described in the paper.

Note that Case 2 is a particular case of Case 3 when consider r3=0. By inspection of Figure 3 in the manuscript, one can see the decrease in variance compared to separate trials with respect to r2, and that the maximum variance reduction occurs for r1+r2=1, and thus for a two-period trial.

### Case 3 (Section 3.3)

**Set and simplify conditions**

In[1]:= `subst = {n11 → r1 ∗ N / 2 , n01 → r1 ∗ N / 2 ,`
   `n12 → r2 ∗ N − n02 − n22 , n03 → r3 ∗ N / 2 , n23 → r3 ∗ N / 2};`
  `substp = {n12 → r2 ∗ N ∗ p12, n22 → r2 ∗ N ∗ p22, n02 → r2 ∗ N ∗ p02, r3 → 1 − r1 − r2};`

**Define terms to optimise.**

Note: *sigma\*term1^(-1)/N* is the variance of the estimator of effect 1 (analogously *sigma\*term2^(-1)/N* for effect 2). But since sigma and N are fixed, we simply work on term1 and term2 expressions.

In[3]:= `term1 =`
  `FullSimplify[((n11 ∗ n01 / (n11 + n01)) + (n12 ∗ n02 / (n12 + n02))) / N /. subst /. substp]`

Out[3]= $\dfrac{r1}{4} + \dfrac{p02\ (-1 + p02 + p22)\ r2}{-1 + p22}$

In[4]:= `term2 =`
  `FullSimplify[(n22 ∗ n02 / (n22 + n02) / N) + (n23 ∗ n03 / (n23 + n03) / N) /. subst /. substp]`

Out[4]= $\dfrac{1}{4}\ (1 - r1 - r2) + \dfrac{p02\ p22\ r2}{p02 + p22}$

In[5]:= `constr = FullSimplify[term1 − term2]`

Out[5]= $\dfrac{1}{4}\left(-1 + 2\ r1 + r2 + \dfrac{4\ p02\ (-1 + p02 + p22)\ r2}{-1 + p22} - \dfrac{4\ p02\ p22\ r2}{p02 + p22}\right)$

In[6]:= **e1 = Solve[D[term2, p02] == l D[constr, p02], l]**
**e2 = Solve[D[term2, p22] == l D[constr, p22], l]**

Out[6]= $\left\{\left\{l \to \dfrac{4\left(\frac{p02\,p22\,r2}{(p02+p22)^2} - \frac{p22\,r2}{p02+p22}\right)}{-\frac{4\,p02\,r2}{-1+p22} - \frac{4\,(-1+p02+p22)\,r2}{-1+p22} - \frac{4\,p02\,p22\,r2}{(p02+p22)^2} + \frac{4\,p22\,r2}{p02+p22}}\right\}\right\}$

Out[7]= $\left\{\left\{l \to \dfrac{4\left(\frac{p02\,p22\,r2}{(p02+p22)^2} - \frac{p02\,r2}{p02+p22}\right)}{-\frac{4\,p02\,r2}{-1+p22} + \frac{4\,p02\,(-1+p02+p22)\,r2}{(-1+p22)^2} - \frac{4\,p02\,p22\,r2}{(p02+p22)^2} + \frac{4\,p02\,r2}{p02+p22}}\right\}\right\}$

In[8]:= **e3 = e1〚1〛〚1〛〚2〛 == e2〚1〛〚1〛〚2〛**

Out[8]= $\dfrac{4\left(\frac{p02\,p22\,r2}{(p02+p22)^2} - \frac{p22\,r2}{p02+p22}\right)}{-\frac{4\,p02\,r2}{-1+p22} - \frac{4\,(-1+p02+p22)\,r2}{-1+p22} - \frac{4\,p02\,p22\,r2}{(p02+p22)^2} + \frac{4\,p22\,r2}{p02+p22}}$ == $\dfrac{4\left(\frac{p02\,p22\,r2}{(p02+p22)^2} - \frac{p02\,r2}{p02+p22}\right)}{-\frac{4\,p02\,r2}{-1+p22} + \frac{4\,p02\,(-1+p02+p22)\,r2}{(-1+p22)^2} - \frac{4\,p02\,p22\,r2}{(p02+p22)^2} + \frac{4\,p02\,r2}{p02+p22}}$

In[9]:= **sol = Solve[e3, {p02}]〚3〛**

Out[9]= $\left\{p02 \to \dfrac{-1 + 2\,p22 - 2\,p22^2}{2\,(-1+p22)}\right\}$

The solution corresponds to the optimal allocation for p02 (Section 3.3 in the paper):

In[10]:= **FullSimplify[sol]**

Out[10]= $\left\{p02 \to \dfrac{1}{2 - 2\,p22} - p22\right\}$

Where the optimal solution for p22 can be obtained numerically by solving the following equation:

In[11]:= **eq2 = Assuming[{r2 + r3 > 1 / 2, r2 + r3 < 1, p22 > 0, p22 < 1},**
**FullSimplify[(term1 - term2) /. sol]]**

Out[11]= $\dfrac{1}{4}\left(-1 + 2\,r1 + \dfrac{(-2 + p22\,(11 + p22\,(-29 + p22\,(49 - 4\,p22\,(13 + 2\,(-4 + p22)\,p22)))))\,r2}{(-1+p22)^3}\right)$

In[12]:= **soleq = Assuming[{p22 > 0, p22 < 1}, FullSimplify[Solve[(eq2 == 0), r2]]]**

Out[12]= $\left\{\left\{r2 \to \dfrac{(-1+p22)^3\,(-1+2\,r1)}{(-1+2\,p22)\,(-2 + p22\,(7 + p22\,(-15 + p22\,(19 + 2\,p22\,(-7 + 2\,p22)))))}\right\}\right\}$

# Supplementary Material of "Optimal allocation strategies in platform trials"

## Design with concurrent and non-concurrent controls - Case 2

In this file, we provide the derivation of the optimal allocation for trials utilising concurrent and non-concurrent controls. Next, we focus on determining the optimal solutions for the Case 2 described in the paper.

For simplicity of calculation, in this file we start optimising using a different parametrization. We consider: $r_{i,s} = n_{i,s} / N_s$, where as before $n_{i,s}$ is the sample size for arm i in the period s, and $N_s$ is the total sample size in period s. At the end of the document we express the solutions in terms of $p_{i,s}$ as in the paper.

**Set conditions**

In[1]:= `subst = {r11 → r1 / 2, r01 → r1 / 2, r02 → (1 - r1) - r12 - r22 };`

In[2]:= `substp = {r12 → r2 p12, r22 → r2 p22, r02 → r2 p02, r3 → 1 - r1 - r2};`

**Define terms to optimise.**

In[3]:= `term1 = FullSimplify[(r11 * r01 / (r11 + r01)) + (r12 * r02 / (r12 + r02)) /. subst]`

Out[3]= $\dfrac{r1}{4} + r12 + \dfrac{r12^2}{-1 + r1 + r22}$

In[4]:= `term2 = FullSimplify[`
    `(1 / r22 + 1 / r02 - ((1 / r02)^2 / (1 / r01 + 1 / r02 + 1 / r11 + 1 / r12))) ^ (-1) /. subst]`

Out[4]= $\dfrac{r22 \left(r1^2 + 4 \, r12 \, (-1 + r12 + r22) + r1 \, (-1 + 4 \, r12 + r22)\right)}{r1^2 + 4 \, (-1 + r12) \, r12 + r1 \, (-1 + 4 \, r12)}$

In[5]:= `sol = Solve[term1 == term2, r12] ⟦3⟧`

Out[5]= $\left\{ r12 \rightarrow \dfrac{1}{2} \left(1 - r1 - \sqrt{1 - r1 - 4 \, r22 + 4 \, r1 \, r22 + 4 \, r22^2}\right) \right\}$

In[6]:= `term3 = Simplify[term1 /. sol]`

Out[6]= $\dfrac{r22 \left(-2 + 3 \, r1 + 4 \, r22 - 2 \sqrt{(1 - 2 \, r22)^2 + r1 \, (-1 + 4 \, r22)}\right)}{4 \, (-1 + r1 + r22)}$

In[7]:= **derivative = FullSimplify[D[term3, r22]]**

Out[7]= $\frac{1}{4\,(-1+r1+r22)^2}\left(r22\,(-1+r1+r22)\left(4-\frac{4\,(-1+r1+2\,r22)}{\sqrt{(1-2\,r22)^2+r1\,(-1+4\,r22)}}\right)-\right.$

$\qquad r22\left(-2+3\,r1+4\,r22-2\,\sqrt{(1-2\,r22)^2+r1\,(-1+4\,r22)}\right)+$

$\qquad (-1+r1+r22)\left(-2+3\,r1+4\,r22-2\,\sqrt{(1-2\,r22)^2+r1\,(-1+4\,r22)}\right)\bigg)$

In[8]:= **sol2 = Solve[derivative == 0, r22];**

⋯ Solve: There may be values of the parameters for which some or all solutions are not valid.

In[9]:= **sol22 = Assuming[{r1 > 0, r1 < 1 / 2}, Simplify[sol2[[3]]]]**

Out[9]= $\left\{r22 \to 1-r1+\frac{1}{4\sqrt{3}}\left(\sqrt{\left(\left((-1+r1)\left(9\,r1^2+6\,r1\right.\right.\right.}\right.\right.$

$\qquad\qquad\left(-4+\left(8+36\,r1-108\,r1^2+27\,r1^3+6\,\sqrt{3}\,\sqrt{r1\left(16-72\,r1+108\,r1^2-27\,r1^3\right)}\right)^{1/3}\right)+$

$\qquad\qquad\left(-2+\left(8+36\,r1-108\,r1^2+27\,r1^3+6\,\sqrt{3}\right.\right.$

$\qquad\qquad\qquad\left.\left.\left.\sqrt{r1\left(16-72\,r1+108\,r1^2-27\,r1^3\right)}\right)^{1/3}\right)^2\right)\right)\bigg/$

$\qquad\quad\left(8+36\,r1-108\,r1^2+27\,r1^3+6\,\sqrt{3}\,\sqrt{r1\left(16-72\,r1+108\,r1^2-27\,r1^3\right)}\right)^{1/3}\Bigg)-$

$\qquad\quad\frac{1}{2}\left(\left(\left(-\frac{22}{3}+8\,(-1+r1)^2+\frac{43\,r1}{3}-7\,r1^2-\right.\right.\right.$

$\qquad\qquad\frac{(-1+r1)\left(4-24\,r1+9\,r1^2\right)}{12\left(8+36\,r1-108\,r1^2+27\,r1^3+6\,\sqrt{3}\,\sqrt{r1\left(16-72\,r1+108\,r1^2-27\,r1^3\right)}\right)^{1/3}}-$

$\qquad\qquad\frac{1}{12}\,(-1+r1)\left(8+36\,r1-108\,r1^2+27\,r1^3+6\,\sqrt{3}\,\sqrt{r1\left(16-72\,r1+108\,r1^2-27\,r1^3\right)}\right)^{1/3}+$

$\qquad\qquad\left(\sqrt{3}\,(-1+r1)^2\,r1\right.$

$\qquad\qquad\qquad\left(8+36\,r1-108\,r1^2+27\,r1^3+6\,\sqrt{3}\,\sqrt{r1\left(16-72\,r1+108\,r1^2-27\,r1^3\right)}\right)^{1/6}\bigg)\bigg/$

$\qquad\qquad\left(\sqrt{\left((-1+r1)\left(9\,r1^2+6\,r1\left(-4+\left(8+36\,r1-108\,r1^2+27\,r1^3+\right.\right.\right.\right.}\right.$

$\qquad\qquad\qquad6\,\sqrt{3}\,\sqrt{r1\left(16-72\,r1+108\,r1^2-27\,r1^3\right)}\Big)^{1/3}\Big)+\Big(-2+\Big(8+36\,r1-$

$\qquad\qquad\qquad108\,r1^2+27\,r1^3+6\,\sqrt{3}\,\sqrt{r1\left(16-72\,r1+108\,r1^2-27\,r1^3\right)}\Big)^{1/3}\Big)^2\Big)\Big)\Big)\Big)\bigg)\bigg\}$

In[10]:= `CForm[sol22[[1]][[2]]]`

Out[10]//CForm=
```
1 - r1 + Sqrt(((-1 + r1)*(9*Power(r1,2) + 6*r1*(-4 + Power(8 + 36*r1 - 108*Power(r1,2)
         0.3333333333333333))) + Power(-2 + Power(8 + 36*r1 - 108*Power(r1,2) + 27*
         0.3333333333333333),2)))/Power(8 + 36*r1 - 108*Power(r1,2) + 27*Power(r1,3
    (4.*Sqrt(3)) - Sqrt(-7.333333333333333 + 8*Power(-1 + r1,2) + (43*r1)/3. - 7*Power
    ((-1 + r1)*(4 - 24*r1 + 9*Power(r1,2)))/
    (12.*Power(8 + 36*r1 - 108*Power(r1,2) + 27*Power(r1,3) + 6*Sqrt(3)*Sqrt(r1*(16
    ((-1 + r1)*Power(8 + 36*r1 - 108*Power(r1,2) + 27*Power(r1,3) + 6*Sqrt(3)*Sqrt(r
    (Sqrt(3)*Power(-1 + r1,2)*r1*Power(8 + 36*r1 - 108*Power(r1,2) + 27*Power(r1,3)
    Sqrt((-1 + r1)*(9*Power(r1,2) + 6*r1*(-4 + Power(8 + 36*r1 - 108*Power(r1,2) +
         0.3333333333333333))) + Power(-2 + Power(8 + 36*r1 - 108*Power(r1,2) + 27
         0.3333333333333333),2)))))/2.
```

In[11]:= `sol22 /. {r1 → 0.3, r2 → 0.7}`

Out[11]= $\{r22 \to 0.302281\}$

Next, we simplify the solution by defining terms a and b.

In[12]:= $\texttt{sol22a = FullSimplify}\Big[\texttt{sol22 /. }\Big\{\texttt{FullSimplify}\Big[$
$\Big(8 + 36\,r1 - 108\,r1^2 + 27\,r1^3 + 6\,\sqrt{3}\,\sqrt{r1\,(16 - 72\,r1 + 108\,r1^2 - 27\,r1^3)}\Big)^{1/3} \to a\Big]\Big\}\Big]$

Out[12]= $\Big\{r22 \to 1 - r1 + \dfrac{1}{4\sqrt{3}} \Big(\sqrt{\Big(\big((-1 + r1)\,\big(9\,r1^2 + 6\,r1\,\big(-4 + \big(8 + 9\,r1\,(4 + 3\,(-4 + r1)\,r1) +}$
$6\,\sqrt{3}\,\sqrt{r1\,(16 - 9\,r1\,(8 + 3\,(-4 + r1)\,r1))}\big)^{1/3}\big) + \big(-2 + \big(8 + 9\,r1\,(4 +$
$3\,(-4 + r1)\,r1) + 6\,\sqrt{3}\,\sqrt{r1\,(16 - 9\,r1\,(8 + 3\,(-4 + r1)\,r1))}\big)^{1/3}\big)^2\big)\Big) \Big/$
$\big(8 + 9\,r1\,(4 + 3\,(-4 + r1)\,r1) + 6\,\sqrt{3}\,\sqrt{r1\,(16 - 9\,r1\,(8 + 3\,(-4 + r1)\,r1))}\big)^{1/3}\Big)\Big) -$
$\dfrac{1}{2}\sqrt{\Bigg(-\dfrac{22}{3} + 8\,(-1 + r1)^2 + \dfrac{43\,r1}{3} - 7\,r1^2 -}$
$\dfrac{(-1 + r1)\,(4 + 3\,r1\,(-8 + 3\,r1))}{12\,\big(8 + 9\,r1\,(4 + 3\,(-4 + r1)\,r1) + 6\,\sqrt{3}\,\sqrt{r1\,(16 - 9\,r1\,(8 + 3\,(-4 + r1)\,r1))}\big)^{1/3}} - \dfrac{1}{12}$
$(-1 + r1)\,\big(8 + 9\,r1\,(4 + 3\,(-4 + r1)\,r1) + 6\,\sqrt{3}\,\sqrt{r1\,(16 - 9\,r1\,(8 + 3\,(-4 + r1)\,r1))}\big)^{1/3} +$
$\Big(\sqrt{3}\,(-1 + r1)^2\,r1$
$\big(8 + 9\,r1\,(4 + 3\,(-4 + r1)\,r1) + 6\,\sqrt{3}\,\sqrt{r1\,(16 - 9\,r1\,(8 + 3\,(-4 + r1)\,r1))}\big)^{1/6} \Big/$
$\Big(\sqrt{\Big((-1 + r1)\,\big(9\,r1^2 + 6\,r1\,\big(-4 + \big(8 + 9\,r1\,(4 + 3\,(-4 + r1)\,r1) + 6\,\sqrt{3}}$
$\sqrt{r1\,(16 - 9\,r1\,(8 + 3\,(-4 + r1)\,r1))}\big)^{1/3}\big) + \big(-2 + \big(8 + 9\,r1\,(4 + 3\,(-4 +}$
$r1)\,r1) + 6\,\sqrt{3}\,\sqrt{r1\,(16 - 9\,r1\,(8 + 3\,(-4 + r1)\,r1))}\big)^{1/3}\big)^2\big)\Big)\Big)\Big)\Big\}$

In[13]:= $\texttt{FullSimplify}\Big[\big(8 + 9\,r1\,(4 + 3\,(-4 + r1)\,r1) + 6\,\sqrt{3}\,\sqrt{r1\,(16 - 9\,r1\,(8 + 3\,(-4 + r1)\,r1))}\big)^{1/3} \Big/$
$\big(8 + 36\,r1 - 108\,r1^2 + 27\,r1^3 + 6\,\sqrt{3}\,\sqrt{r1\,(16 - 72\,r1 + 108\,r1^2 - 27\,r1^3)}\big)^{1/3}\Big]$

Out[13]= 1

In[14]:= **sol22a** /. $\left\{ \left(8 + 9\,r1\,(4 + 3\,(-4 + r1)\,r1) + 6\,\sqrt{3}\,\sqrt{r1\,(16 - 9\,r1\,(8 + 3\,(-4 + r1)\,r1))} \right)^{1/3} \to a \right\}$

Out[14]= $\left\{ r22 \to 1 - r1 + \dfrac{\sqrt{\dfrac{(-1 + r1)\,\left((-2 + a)^2 + 6\,(-4 + a)\,r1 + 9\,r1^2\right)}{\left(8 + 9\,r1\,(4 + 3\,(-4 + r1)\,r1) + 6\,\sqrt{3}\,\sqrt{r1\,(16 - 9\,r1\,(8 + 3\,(-4 + r1)\,r1))}\right)^{1/3}}}}{4\,\sqrt{3}} - \right.$

$\dfrac{1}{2}\,\sqrt{\left(-\dfrac{22}{3} - \dfrac{1}{12}\,a\,(-1 + r1) + 8\,(-1 + r1)^2 + \dfrac{43\,r1}{3} - 7\,r1^2 - \right.}$

$\dfrac{(-1 + r1)\,(4 + 3\,r1\,(-8 + 3\,r1))}{12\,\left(8 + 9\,r1\,(4 + 3\,(-4 + r1)\,r1) + 6\,\sqrt{3}\,\sqrt{r1\,(16 - 9\,r1\,(8 + 3\,(-4 + r1)\,r1))}\right)^{1/3}} +$

$\left(\sqrt{3}\,(-1 + r1)^2\,r1 \right.$

$\left(8 + 9\,r1\,(4 + 3\,(-4 + r1)\,r1) + 6\,\sqrt{3}\,\sqrt{r1\,(16 - 9\,r1\,(8 + 3\,(-4 + r1)\,r1))}\right)^{1/6}\Big/$

$\left.\left.\left.\left(\sqrt{(-1 + r1)\,\left((-2 + a)^2 + 6\,(-4 + a)\,r1 + 9\,r1^2\right)}\right)\right)\right\}$

In[15]:= **r22a = 1 - r1 + (√(((-1 + r1) ((-2 + a)^2 + 6 (-4 + a) r1 + 9 r1^2)) / a)) / (4 × √3) -**
**(1 / 2) √(-(22 / 3) - (1 / 12) a (-1 + r1) + 8 (-1 + r1)^2 +**
**(43 r1) / 3 - 7 r1^2 - ((-1 + r1) (4 + 3 r1 (-8 + 3 r1))) / (12 a) +**
**(√3 (-1 + r1)^2 r1 (a^3)^(1/6)) / (√((-1 + r1) ((-2 + a)^2 + 6 (-4 + a) r1 + 9 r1^2))))**

Out[15]= $1 - r1 + \dfrac{\sqrt{\dfrac{(-1 + r1)\,\left((-2 + a)^2 + 6\,(-4 + a)\,r1 + 9\,r1^2\right)}{a}}}{4\,\sqrt{3}} -$

$\dfrac{1}{2}\,\sqrt{\left(-\dfrac{22}{3} - \dfrac{1}{12}\,a\,(-1 + r1) + 8\,(-1 + r1)^2 + \dfrac{43\,r1}{3} - 7\,r1^2 + \right.}$

$\left. \dfrac{\sqrt{3}\,(a^3)^{1/6}\,(-1 + r1)^2\,r1}{\sqrt{(-1 + r1)\,\left((-2 + a)^2 + 6\,(-4 + a)\,r1 + 9\,r1^2\right)}} - \dfrac{(-1 + r1)\,(4 + 3\,r1\,(-8 + 3\,r1))}{12\,a} \right)$

In[16]:= **CForm[r22a]**

Out[16]//CForm=
```
1 - r1 + Sqrt(((-1 + r1)*(Power(-2 + a,2) + 6*(-4 + a)*r1 + 9*Power(r1,2)))/a)/(4.*Sq
    Sqrt(-7.333333333333333 - (a*(-1 + r1))/12. + 8*Power(-1 + r1,2) + (43*r1)/3. - 7*P
     (Sqrt(3)*Power(Power(a,3),0.16666666666666666)*Power(-1 + r1,2)*r1)/Sqrt((-1 + r
    2.
```

In[17]:= **FullSimplify[r22a]**

Out[17]= $1 - r1 + \dfrac{\sqrt{\dfrac{(-1+r1)\ ((-2+a)^2 + 6\ (-4+a)\ r1 + 9\ r1^2)}{a}}}{4\ \sqrt{3}}\ -$

$\dfrac{1}{2}\ \sqrt{\left(-\dfrac{22}{3} - \dfrac{1}{12}\ a\ (-1+r1) + 8\ (-1+r1)^2 + \dfrac{43\ r1}{3} - 7\ r1^2 + \dfrac{\sqrt{3}\ (a^3)^{1/6}\ (-1+r1)^2\ r1}{\sqrt{(-1+r1)\ ((-2+a)^2 + 6\ (-4+a)\ r1 + 9\ r1^2)}} - \dfrac{(-1+r1)\ (4 + 3\ r1\ (-8 + 3\ r1))}{12\ a}\right)}$

In[18]:= **sol22ab = FullSimplify$\left[$r22a /. $\left\{(-2+a)^2 + 6\ (-4+a)\ r1 + 9\ r1^2 \rightarrow b\right\}\right]$**

Out[18]= $\dfrac{1}{12}\left(12 + \sqrt{3}\ \sqrt{\dfrac{b\ (-1+r1)}{a}} - 12\ r1 - \sqrt{3}\ \sqrt{-\dfrac{(-1+r1)\ \left(-12\ \sqrt{3}\ a\ (a^3)^{1/6}\ \sqrt{b\ (-1+r1)}\ r1 + b\ (4 + 8\ a + a^2 - 12\ (2+a)\ r1 + 9\ r1^2)\right)}{a\ b}}\right)$

In[19]:= **sol22abs = FullSimplify$\left[$PowerExpand$\left[$r22a /. $\left\{(-2+a)^2 + 6\ (-4+a)\ r1 + 9\ r1^2 \rightarrow b\right\}\right]\right]$**

Out[19]= $\dfrac{1}{12}\left(12 + \dfrac{\sqrt{3}\ \sqrt{b}\ \sqrt{-1+r1}}{\sqrt{a}} - 12\ r1 - \sqrt{3}\ \sqrt{-\dfrac{(-1+r1)\ \left(-12\ \sqrt{3}\ a^{3/2}\ \sqrt{-1+r1}\ r1 + \sqrt{b}\ (4 + 8\ a + a^2 - 12\ (2+a)\ r1 + 9\ r1^2)\right)}{a\ \sqrt{b}}}\right)$

In[20]:= **CForm[sol22ab]**

Out[20]//CForm=
```
(12 + Sqrt(3)*Sqrt((b*(-1 + r1))/a) - 12*r1 - Sqrt(3)*Sqrt(-(((-1 + r1)*(-12*Sqrt(3)*
     b*(4 + 8*a + Power(a,2) - 12*(2 + a)*r1 + 9*Power(r1,2))))/(a*b)))))/12.
```

We compute the solution in terms of p22 by taking into account that p22=r22/(1-r1), as 1-r1=r2 in this case.

In[21]:= **solp22 = FullSimplify[sol22ab / (1 - r1)]**

Out[21]= $\dfrac{-12 - \sqrt{3}\ \sqrt{\dfrac{b\ (-1+r1)}{a}} + 12\ r1 + \sqrt{3}\ \sqrt{-\dfrac{(-1+r1)\ \left(-12\ \sqrt{3}\ a\ (a^3)^{1/6}\ \sqrt{b\ (-1+r1)}\ r1 + b\ (4 + 8\ a + a^2 - 12\ (2+a)\ r1 + 9\ r1^2)\right)}{a\ b}}}{12\ (-1+r1)}$

In[22]:= **CForm[solp22]**

Out[22]//CForm=
```
(-12 - Sqrt(3)*Sqrt((b*(-1 + r1))/a) + 12*r1 + Sqrt(3)*Sqrt(-(((-1 + r1)*(-12*Sqrt(3)*
     b*(4 + 8*a + Power(a,2) - 12*(2 + a)*r1 + 9*Power(r1,2))))/(a*b))))/(12
```

### Simplifying solution p22

First note that a>0 and b<0

In[23]:= $\left(8 + 9 \; r1 \; (4 + 3 \; (-4 + r1) \; r1) + 6 \; \sqrt{3} \; \sqrt{r1 \; (16 - 9 \; r1 \; (8 + 3 \; (-4 + r1) \; r1))}\right)^{1/3}$ /. {r1 → 0.3}

Out[23]= 2.72751

In[24]:= $(-2 + a)^2 + 6 \; (-4 + a) \; r1 + 9 \; r1^2$ /. {r1 → 0.3, a → 2.727510816380422}

Out[24]= -0.951209

In[25]:= FullSimplify$\left[\dfrac{-12 + 12 \; r1}{12 \; (-1 + r1)}\right]$ + FullSimplify$\Big[$

$\dfrac{-\sqrt{3} \; \sqrt{b \; (-1 + r1) \, / \, a} \; + \sqrt{3} \; \sqrt{-\dfrac{(-1+r1) \; \left(-12 \; \sqrt{3} \; a \; \left(a^3\right)^{1/6} \; \sqrt{b \; (-1+r1)} \; r1 + b \; \left(4 + 8 \; a + a^2 - 12 \; (2+a) \; r1 + 9 \; r1^2\right)\right)}{a \, b}}}{12 \; (-1 + r1)}\Big]$

Out[25]= $1 + \dfrac{-\sqrt{\dfrac{b \; (-1+r1)}{a}} + \sqrt{-\dfrac{(-1+r1) \; \left(-12 \; \sqrt{3} \; a \; \left(a^3\right)^{1/6} \; \sqrt{b \; (-1+r1)} \; r1 + b \; \left(4 + 8 \; a + a^2 - 12 \; (2+a) \; r1 + 9 \; r1^2\right)\right)}{a \, b}}}{4 \; \sqrt{3} \; (-1 + r1)}$

In[26]:= solp22s = Assuming$\Big[$ {a > 0, b < 0, r1 < 1, r1 > 0},

FullSimplify$\Big[1 + \dfrac{-\sqrt{\dfrac{b \; (-1+r1)}{a}} + \sqrt{-\dfrac{(-1+r1) \; \left(-12 \; \sqrt{3} \; a \; \left(a^3\right)^{1/6} \; \sqrt{b \; (-1+r1)} \; r1 + b \; \left(4 + 8 \; a + a^2 - 12 \; (2+a) \; r1 + 9 \; r1^2\right)\right)}{a \, b}}}{4 \; \sqrt{3} \; (-1 + r1)}\Big]\Big]$

Out[26]= $1 + \dfrac{-\sqrt{\dfrac{b \; (-1+r1)}{a}} + \sqrt{-\dfrac{(-1+r1) \; \left(-12 \; \sqrt{3} \; \sqrt{a^3 \, b} \; (-1+r1) \; r1 + b \; \left(4 + 8 \; a + a^2 - 12 \; (2+a) \; r1 + 9 \; r1^2\right)\right)}{a \, b}}}{4 \; \sqrt{3} \; (-1 + r1)}$

In[27]:= solp22s2 = $1 + \dfrac{\sqrt{-b} \; - \sqrt{12 \; \sqrt{3} \; \sqrt{a^3 \; b^{\wedge}(-1)} \; (-1+r1) \; r1 + \left(4 + 8 \; a + a^2 - 12 \; (2+a) \; r1 + 9 \; r1^2\right)}}{4 \; \sqrt{3 \; a} \; (1 - r1)}$

Out[27]= $1 + \dfrac{\sqrt{-b} \; - \sqrt{4 + 8 \; a + a^2 - 12 \; (2+a) \; r1 + 12 \; \sqrt{3} \; \sqrt{\dfrac{a^3 \; (-1+r1)}{b}} \; r1 + 9 \; r1^2}}{4 \; \sqrt{3} \; \sqrt{a \; (1-r1)}}$

In[28]:= CForm[solp22s2]

Out[28]//CForm=
1 + (Sqrt(-b) - Sqrt(4 + 8*a + Power(a,2) - 12*(2 + a)*r1 + 12*Sqrt(3)*Sqrt((Power(a,3

In[29]:= solp22s /. {r1 → 0.3, a → 2.727510816380422`, b → -0.9512085425647322`}

Out[29]= 0.43183

In[30]:= solp22s2 /. {r1 → 0.3, a → 2.727510816380422`, b → -0.9512085425647322`}

Out[30]= 0.43183

In[31]:= Assuming[ {a > 0, b < 0, r1 < 1, r1 > 0}, FullSimplify[solp22s / solp22]]

Out[31]= 1

Finally, to get the solution r12 (and hence p12), one needs to use the solution obtained for r22 and substitute it in the expression for r12 (**sol**).

In[32]:= **sol**

Out[32]= $\left\{ r12 \to \frac{1}{2} \left( 1 - r1 - \sqrt{1 - r1 - 4\ r22 + 4\ r1\ r22 + 4\ r22^2} \right) \right\}$

In[33]:= **solp22 = FullSimplify[sol[[1]][[2]] / (1 - r1) /. {r22 → p22 ∗ (1 - r1)}]**

Out[33]= $\dfrac{-1 + \sqrt{(-1 + 4\ (-1 + p22)\ p22\ (-1 + r1))\ (-1 + r1)} + r1}{2\ (-1 + r1)}$

In[34]:= **solp22s = FullSimplify**$\Big[$**solp22 /.**

$\left\{ p22 \to 1 + \dfrac{\sqrt{-b\ (1 - r1)} - \sqrt{\frac{(1-r1)\ \left(-12\ \sqrt{3}\ a\ (a^3)^{1/6}\ \sqrt{-b\ (1-r1)}\ r1 + b\ \left(4 + 8\ a + a^2 - 12\ (2+a)\ r1 + 9\ r1^2\right)\right)}{b}}}{4\ \sqrt{3\ a}\ (1 - r1)} \right\}\Big]$

Out[34]= $\dfrac{1}{2} + \dfrac{1}{2\ (-1 + r1)}$

$\left( \sqrt{\left( (-1 + r1)\ \left( -1 + \dfrac{1}{12\ \sqrt{3}\ a\ (-1 + r1)} \left( \sqrt{b\ (-1 + r1)} - \sqrt{\left( -\dfrac{1}{b}\ (-1 + r1)\ \left( -12\ \sqrt{3}\ a \right.\right.} \right.\right.\right.}\right.$

$\left. \left. \left. \left. (a^3)^{1/6}\ \sqrt{b\ (-1 + r1)}\ r1 + b\ \left( 4 + 8\ a + a^2 - 12\ (2 + a)\ r1 + 9\ r1^2 \right) \right) \right) \right) \right)$

$\left( -12\ \sqrt{a}\ (-1 + r1) + \sqrt{3} \left( \sqrt{b\ (-1 + r1)} - \sqrt{\left( -\dfrac{1}{b}\ (-1 + r1)\ \left( -12\ \sqrt{3}\ a\ (a^3)^{1/6} \right.\right.} \right. \right.$

$\left. \left. \left. \left. \left. \left. \sqrt{b\ (-1 + r1)}\ r1 + b\ \left( 4 + 8\ a + a^2 - 12\ (2 + a)\ r1 + 9\ r1^2 \right) \right) \right) \right) \right) \right) \right)$

# Supplementary Material of "Optimal allocation strategies in platform trials"

## Design with concurrent and non-concurrent controls - Case 3

We first define the treatment effect for arm 2 following the expressions presented in the supplementary material (see Section A.2). To do so, we define the matrices A, B, and C, and use equation (1.b) to to obtain point estimates

```
In[∘]:=  nd1 = n01 + n11
         nd2 = n02 + n12 + n22
         nd3 = n03 + n23
         A = {{nd1, 0, 0}, {0, nd2, 0}, {0, 0, nd3}}
         B = {{n11, 0}, {n12, n22}, {0, n23}}
         Cm = {{n11 + n12, 0}, {0, n22 + n23}}
```

Out[∘]=  n01 + n11

Out[∘]=  n02 + n12 + n22

Out[∘]=  n03 + n23

Out[∘]=  {{n01 + n11, 0, 0}, {0, n02 + n12 + n22, 0}, {0, 0, n03 + n23}}

Out[∘]=  {{n11, 0}, {n12, n22}, {0, n23}}

Out[∘]=  {{n11 + n12, 0}, {0, n22 + n23}}

*In[ ]:=* `M = FullSimplify[Inverse[Cm - Transpose[B].Inverse[A].B]]`
`Nm = {{n11 * theta11 + n12 * theta12}, {n22 * theta22 + n23 * theta23}}`
`Collect[FullSimplify[M.Nm][[2]], {theta11, theta12, theta22, theta23}]`
`w11 = (n11 (n01 + n11) n12 n22 (n03 + n23) ) /`
`   (n11 n12 (n02 n03 n22 + n03 n22 n23 + n02 (n03 + n22) n23) +`
`     n01 (n03 n11 n12 n22 + (n03 n11 n12 + n11 n12 n22 + n03 (n11 + n12) n22) n23 +`
`       n02 (n11 + n12) (n22 n23 + n03 (n22 + n23))))`
`w12 = ((n01 + n11) n12^2 n22 (n03 + n23) ) /`
`   (n11 n12 (n02 n03 n22 + n03 n22 n23 + n02 (n03 + n22) n23) +`
`     n01 (n03 n11 n12 n22 + (n03 n11 n12 + n11 n12 n22 + n03 (n11 + n12) n22) n23 +`
`       n02 (n11 + n12) (n22 n23 + n03 (n22 + n23))))`
`w22 =`
`   (n22 (n11 n12 (n02 + n22) + n01 (n11 n12 + n02 (n11 + n12) + (n11 + n12) n22)) (n03 + n23) ) /`
`   (n11 n12 (n02 n03 n22 + n03 n22 n23 + n02 (n03 + n22) n23) +`
`     n01 (n03 n11 n12 n22 + (n03 n11 n12 + n11 n12 n22 + n03 (n11 + n12) n22) n23 +`
`       n02 (n11 + n12) (n22 n23 + n03 (n22 + n23))))`
`w23 =`
`   ((n11 n12 (n02 + n22) + n01 (n11 n12 + n02 (n11 + n12) + (n11 + n12) n22)) n23 (n03 + n23) ) /`
`   (n11 n12 (n02 n03 n22 + n03 n22 n23 + n02 (n03 + n22) n23) +`
`     n01 (n03 n11 n12 n22 + (n03 n11 n12 + n11 n12 n22 + n03 (n11 + n12) n22) n23 +`
`       n02 (n11 + n12) (n22 n23 + n03 (n22 + n23))))`
`sol = M.Nm`
`True ===`
`  FullSimplify[sol[[2]][[1]] == w11 * theta11 + w12 * theta12 + w22 * theta22 + w23 * theta23]`

*Out[ ]=* `{{((n01 + n11) (n03 (n02 + n12) n22 + n03 (n02 + n12) n23 + (n02 + n03 + n12) n22 n23)) /`
`    (n01 n03 (n11 n12 + n02 (n11 + n12)) n22 +`
`      n01 (n03 n11 n12 + n11 n12 n22 + n03 (n11 + n12) n22 + n02 (n11 + n12) (n03 + n22)) n23 +`
`      n11 n12 (n02 n03 n22 + n03 n22 n23 + n02 (n03 + n22) n23)),`
`   ((n01 + n11) n12 n22 (n03 + n23)) / (n11 n12 (n02 n03 n22 + n03 n22 n23 + n02 (n03 + n22) n23) +`
`      n01 (n03 n11 n12 n22 + n11 n12 n22 + n03 (n11 + n12) n22) n23 +`
`      n02 (n11 + n12) (n22 n23 + n03 (n22 + n23)))},`
`  {((n01 + n11) n12 n22 (n03 + n23)) / (n11 n12 (n02 n03 n22 + n03 n22 n23 + n02 (n03 + n22) n23) +`
`      n01 (n03 n11 n12 n22 + (n03 n11 n12 + n11 n12 n22 + n03 (n11 + n12) n22) n23 +`
`      n02 (n11 + n12) (n22 n23 + n03 (n22 + n23)))),`
`   ((n11 n12 (n02 + n22) + n01 (n11 n12 + n02 (n11 + n12) + (n11 + n12) n22)) (n03 + n23)) /`
`    (n11 n12 (n02 n03 n22 + n03 n22 n23 + n02 (n03 + n22) n23) +`
`      n01 (n03 n11 n12 n22 + (n03 n11 n12 + n11 n12 n22 + n03 (n11 + n12) n22) n23 +`
`      n02 (n11 + n12) (n22 n23 + n03 (n22 + n23))))}}`

*Out[ ]=* `{{n11 theta11 + n12 theta12}, {n22 theta22 + n23 theta23}}`

*Out[∘]=* $\Big\{$ (n11 (n01 + n11) n12 n22 (n03 + n23) theta11) /

(n11 n12 (n02 n03 n22 + n03 n22 n23 + n02 (n03 + n22) n23) +

n01 (n03 n11 n12 n22 + (n03 n11 n12 + n11 n12 n22 + n03 (n11 + n12) n22) n23 +

n02 (n11 + n12) (n22 n23 + n03 (n22 + n23)))) +

$\big($(n01 + n11) n12$^2$ n22 (n03 + n23) theta12$\big)$ /

(n11 n12 (n02 n03 n22 + n03 n22 n23 + n02 (n03 + n22) n23) +

n01 (n03 n11 n12 n22 + (n03 n11 n12 + n11 n12 n22 + n03 (n11 + n12) n22) n23 +

n02 (n11 + n12) (n22 n23 + n03 (n22 + n23)))) +

(n22 (n11 n12 (n02 + n22) + n01 (n11 n12 + n02 (n11 + n12) + (n11 + n12) n22))

(n03 + n23) theta22) / (n11 n12 (n02 n03 n22 + n03 n22 n23 + n02 (n03 + n22) n23) +

n01 (n03 n11 n12 n22 + (n03 n11 n12 + n11 n12 n22 + n03 (n11 + n12) n22) n23 +

n02 (n11 + n12) (n22 n23 + n03 (n22 + n23)))) +

((n11 n12 (n02 + n22) + n01 (n11 n12 + n02 (n11 + n12) + (n11 + n12) n22))

n23 (n03 + n23) theta23) /

(n11 n12 (n02 n03 n22 + n03 n22 n23 + n02 (n03 + n22) n23) +

n01 (n03 n11 n12 n22 + (n03 n11 n12 + n11 n12 n22 + n03 (n11 + n12) n22) n23 +

n02 (n11 + n12) (n22 n23 + n03 (n22 + n23))))$\Big\}$

*Out[∘]=* (n11 (n01 + n11) n12 n22 (n03 + n23)) /

(n11 n12 (n02 n03 n22 + n03 n22 n23 + n02 (n03 + n22) n23) +

n01 (n03 n11 n12 n22 + (n03 n11 n12 + n11 n12 n22 + n03 (n11 + n12) n22) n23 +

n02 (n11 + n12) (n22 n23 + n03 (n22 + n23))))

*Out[∘]=* $\big($(n01 + n11) n12$^2$ n22 (n03 + n23)$\big)$ / (n11 n12 (n02 n03 n22 + n03 n22 n23 + n02 (n03 + n22) n23) +

n01 (n03 n11 n12 n22 + (n03 n11 n12 + n11 n12 n22 + n03 (n11 + n12) n22) n23 +

n02 (n11 + n12) (n22 n23 + n03 (n22 + n23))))

*Out[∘]=* (n22 (n11 n12 (n02 + n22) + n01 (n11 n12 + n02 (n11 + n12) + (n11 + n12) n22)) (n03 + n23)) /

(n11 n12 (n02 n03 n22 + n03 n22 n23 + n02 (n03 + n22) n23) +

n01 (n03 n11 n12 n22 + (n03 n11 n12 + n11 n12 n22 + n03 (n11 + n12) n22) n23 +

n02 (n11 + n12) (n22 n23 + n03 (n22 + n23))))

*Out[∘]=* ((n11 n12 (n02 + n22) + n01 (n11 n12 + n02 (n11 + n12) + (n11 + n12) n22)) n23 (n03 + n23)) /

(n11 n12 (n02 n03 n22 + n03 n22 n23 + n02 (n03 + n22) n23) +

n01 (n03 n11 n12 n22 + (n03 n11 n12 + n11 n12 n22 + n03 (n11 + n12) n22) n23 +

n02 (n11 + n12) (n22 n23 + n03 (n22 + n23))))

*Out[∘]=* {{((n01 + n11) (n03 (n02 + n12) n22 + n03 (n02 + n12) n23 + (n02 + n03 + n12) n22 n23)
    (n11 theta11 + n12 theta12)) / (n01 n03 (n11 n12 + n02 (n11 + n12)) n22 +
    n01 (n03 n11 n12 + n11 n12 n22 + n03 (n11 + n12) n22 + n02 (n11 + n12) (n03 + n22)) n23 +
    n11 n12 (n02 n03 n22 + n03 n22 n23 + n02 (n03 + n22) n23)) +
  ((n01 + n11) n12 n22 (n03 + n23) (n22 theta22 + n23 theta23)) /
   (n11 n12 (n02 n03 n22 + n03 n22 n23 + n02 (n03 + n22) n23) +
    n01 (n03 n11 n12 n22 + (n03 n11 n12 + n11 n12 n22 + n03 (n11 + n12) n22) n23 +
       n02 (n11 + n12) (n22 n23 + n03 (n22 + n23))))},
 {((n01 + n11) n12 n22 (n03 + n23) (n11 theta11 + n12 theta12)) /
   (n11 n12 (n02 n03 n22 + n03 n22 n23 + n02 (n03 + n22) n23) +
    n01 (n03 n11 n12 n22 + (n03 n11 n12 + n11 n12 n22 + n03 (n11 + n12) n22) n23 +
       n02 (n11 + n12) (n22 n23 + n03 (n22 + n23)))) +
  ((n11 n12 (n02 + n22) + n01 (n11 n12 + n02 (n11 + n12) + (n11 + n12) n22))
     (n03 + n23) (n22 theta22 + n23 theta23)) /
   (n11 n12 (n02 n03 n22 + n03 n22 n23 + n02 (n03 + n22) n23) +
    n01 (n03 n11 n12 n22 + (n03 n11 n12 + n11 n12 n22 + n03 (n11 + n12) n22) n23 +
       n02 (n11 + n12) (n22 n23 + n03 (n22 + n23))))}}

*Out[∘]=* True

# Variance computation

To compute the variance of treatment effect 2, first note
Var(theta2) = Var(w11*theta11+w12*theta12+w22*theta22+w23*theta23)

```
In[ ]:= theta11 = n01 / (n01 + n11) * (y11 - y01);
        theta12 =  (n02 + n22) / (n02 + n12 + n22) * y12 -
            ((n02 / (n02 + n12 + n22)) * y02 + (n22 / (n02 + n12 + n22)) * y22);
        theta22 =  (n02 + n12) / (n02 + n12 + n22) * y22 -
            ((n02 / (n02 + n12 + n22)) * y02 + (n12 / (n02 + n12 + n22)) * y12);
        theta23 = n03 / (n03 + n23) * (y23 - y03);
        expr = w11 * theta11 + w12 * theta12 + w22 * theta22 + w23 * theta23;
        Collect[FullSimplify[expr], {y01, y11, y02, y12, y22, y03, y23}];
        expr01 = FullSimplify[
            (-n01 n03 n11 n12 n22 - n01 n11 n12 n22 n23) / (n01 n03 (n11 n12 + n02 (n11 + n12)) n22 +
                n01 (n03 n11 n12 + n11 n12 n22 + n03 (n11 + n12) n22 + n02 (n11 + n12) (n03 + n22)) n23 +
                n11 n12 (n02 n03 n22 + n03 n22 n23 + n02 (n03 + n22) n23))];
        expr02 = FullSimplify[
            (-n01 n02 n03 n11 n22 - n01 n02 n03 n12 n22 - n02 n03 n11 n12 n22 - n01 n02 n11 n22 n23 -
                n01 n02 n12 n22 n23 - n02 n11 n12 n22 n23) / (n01 n03 (n11 n12 + n02 (n11 + n12)) n22 +
                n01 (n03 n11 n12 + n11 n12 n22 + n03 (n11 + n12) n22 + n02 (n11 + n12) (n03 + n22)) n23 +
                n11 n12 (n02 n03 n22 + n03 n22 n23 + n02 (n03 + n22) n23))];
        expr03 = FullSimplify[
            (-n01 n03 n11 n12 n23 - n02 n03 n11 n12 n23 - n03 n11 n12 n22 n23 - n01 n03 n11 (n02 + n22)
                 n23 - n01 n03 n12 (n02 + n22) n23) / (n01 n03 (n11 n12 + n02 (n11 + n12)) n22 +
                n01 (n03 n11 n12 + n11 n12 n22 + n03 (n11 + n12) n22 + n02 (n11 + n12) (n03 + n22)) n23 +
                n11 n12 (n02 n03 n22 + n03 n22 n23 + n02 (n03 + n22) n23))];
        expr11 = FullSimplify[
            (n01 n03 n11 n12 n22 + n01 n11 n12 n22 n23) / (n01 n03 (n11 n12 + n02 (n11 + n12)) n22 +
                n01 (n03 n11 n12 + n11 n12 n22 + n03 (n11 + n12) n22 + n02 (n11 + n12) (n03 + n22)) n23 +
                n11 n12 (n02 n03 n22 + n03 n22 n23 + n02 (n03 + n22) n23))];
        expr12 = FullSimplify[
            (-n01 n03 n11 n12 n22 - n01 n11 n12 n22 n23) / (n01 n03 (n11 n12 + n02 (n11 + n12)) n22 +
                n01 (n03 n11 n12 + n11 n12 n22 + n03 (n11 + n12) n22 + n02 (n11 + n12) (n03 + n22)) n23 +
                n11 n12 (n02 n03 n22 + n03 n22 n23 + n02 (n03 + n22) n23))];
        expr22 = FullSimplify[
            (n02 n03 n11 n12 n22 + n01 n03 (n11 n12 + n02 (n11 + n12)) n22 + n01 n02 n12 n22 n23 + n02 n11
                 n12 n22 n23 + n01 n11 (n02 + n12) n22 n23) / (n01 n03 (n11 n12 + n02 (n11 + n12)) n22 +
                n01 (n03 n11 n12 + n11 n12 n22 + n03 (n11 + n12) n22 + n02 (n11 + n12) (n03 + n22)) n23 +
                n11 n12 (n02 n03 n22 + n03 n22 n23 + n02 (n03 + n22) n23))];
        expr23 = FullSimplify[ (n02 n03 n11 n12 n23 + n03 n11 n12 n22 n23 +
                n01 n03 (n11 n12 + n02 (n11 + n12) + (n11 + n12) n22) n23) /
             (n01 n03 (n11 n12 + n02 (n11 + n12)) n22 +
                n01 (n03 n11 n12 + n11 n12 n22 + n03 (n11 + n12) n22 + n02 (n11 + n12) (n03 + n22)) n23 +
                n11 n12 (n02 n03 n22 + n03 n22 n23 + n02 (n03 + n22) n23))];
        FullSimplify[
          Collect[FullSimplify[expr], {y01, y11, y02, y12, y22, y03, y23}] == expr01 * y01 +
            expr02 * y02 + expr03 * y03 + expr11 * y11 + expr12 * y12 + expr22 * y22 + expr23 * y23];
```

**Variance expression is then term2s\*sigma^2/N, where**

*In[ ]:=* `term2f = FullSimplify[expr01^2 * y01 + expr02^2 * y02 + expr03^2 * y03 + expr11^2 * y11 +`
`    expr12^2 * y12 + expr22^2 * y22 + expr23^2 * y23 /. {y01 → 1 / n01, y02 → 1 / n02,`
`    y03 → 1 / n03, y11 → 1 / n11, y12 → 1 / n12, y22 → 1 / n22, y23 → 1 / n23}];`
`term2s =`
`    FullSimplify[term2f /. {n01 → r01 * Nt, n02 → r02 * Nt, n03 → r03 * Nt, n11 → r11 * Nt,`
`    n12 → r12 * Nt, n22 → r22 * Nt, n23 → r23 * Nt}] * Nt;`

### Define terms to optimise

*In[ ]:=* `substp =`
`    {r11 → r1 p11, r12 → r2 p12, r22 → r2 p22, r02 → r2 p02, r23 → r3 p23, r3 → 1 − r1 − r2};`

*In[ ]:=* `subst = {r11 → r1 / 2, r01 → r1 / 2, r23 → r3 / 2, r03 → r3 / 2, r02 → r2 − r12 − r22 };`
`term1 = FullSimplify[(r11 * r01 / (r11 + r01)) + (r12 * r02 / (r12 + r02)) /. subst /. substp]`
`term2 = FullSimplify[(1 / term2s) /. subst /. substp]`

*Out[ ]=* $\dfrac{r1}{4} + \dfrac{p12\,(-1 + p12 + p22)\,r2}{-1 + p22}$

*Out[ ]=* $\left(r1^2 + r1\left(-1 + \left(-4\,(-1 + p12)\,p12 + (1 - 2\,p22)^2\right)r2\right) + 4\,p12\,r2\right.$
$\left. (-1 + p12 + r2 + 4\,(-1 + p22)\,p22\,r2 + p12\,(-1 + 4\,p22)\,r2)\right) / (-4\,r1 + 16\,(-1 + p12)\,p12\,r2)$

*In[ ]:=* `substg = {r01 → r1 − r11, r03 → r3 − r23, r02 → r2 − r12 − r22 };`
`termg1 =`
`    FullSimplify[(r11 * r01 / (r11 + r01)) + (r12 * r02 / (r12 + r02)) /. substg /. substp]`
`termg2 = FullSimplify[(1 / term2s) /. substg /. substp]`

*Out[ ]=* $p11\,(r1 - p11\,r1) + \dfrac{p12\,(-1 + p12 + p22)\,r2}{-1 + p22}$

*Out[ ]=* $\dfrac{p22\,r2\,(-\,(\,(-1 + p11)\,p11\,(-1 + p22)\,r1) + p12\,(-1 + p12 + p22)\,r2)}{(-1 + p11)\,p11\,r1 + (-1 + p12)\,p12\,r2} + p23\,r3\left(1 + \dfrac{p23\,r3}{-1 + r1 + r2}\right)$

### Numerical example: optimisation assuming balanced design in periods 1 and 3

*In[ ]:=* `ex = {r1 → 0.1, r2 → 0.8, r3 → 0.1};`
`FindMinimum[{(-term1) /. ex, term1 == term2 /. ex, p12 > 0, p22 > 0},`
`    {{p12, r2 / 3 /. ex}, {p22, r2 / 3 /. ex}}]`

*Out[ ]=* `{-0.164091, {p12 → 0.303787, p22 → 0.289682}}`

### Optimisation (approach 1) - here we do not assume balanced design in periods 1 and 3 and thus also allocation rates in periods 1 and 3 are optimized

*In[ ]:=* `ex = {r1 → 0.4, r2 → 0.4, r3 → 0.2};`
`FindMinimum[{(-termg1) /. ex, termg1 == termg2 /. ex,`
`    p12 > 0, p22 > 0, p11 > 0, p23 > 0, p11 < 1, p23 < 1},`
`    {{p11, r1 / 2 /. ex}, {p12, r2 / 3 /. ex}, {p22, r2 / 3 /. ex}, {p23, r3 / 2 /. ex}}]`

*Out[ ]=* `{-0.144071, {p11 → 0.5, p12 → 0.153829, p22 → 0.457912, p23 → 0.5}}`

*In[ ]:=* `FindMinimum[`
    `{-termg1 /. ex, termg1 == termg2 /. ex, p12 > 0, p22 > 0, p11 > 0, p23 > 0, p11 < 1, p23 < 1},`
    `{{p11,` $\frac{r1}{2}$ `/. ex}, {p12,` $\frac{r2}{3}$ `/. ex}, {p22,` $\frac{r2}{3}$ `/. ex}, {p23,` $\frac{r3}{2}$ `/. ex}}]`

*Out[ ]=* `{-0.144071, {p11 → 0.5, p12 → 0.153829, p22 → 0.457912, p23 → 0.5}}`

*In[ ]:=* `termg2`

*Out[ ]=* $\frac{p22\ r2\ (-\ (\ (-1+p11)\ p11\ (-1+p22)\ r1)\ +\ p12\ (-1+p12+p22)\ r2)}{(-1+p11)\ p11\ r1\ +\ (-1+p12)\ p12\ r2}\ +\ p23\ r3\ \left(1+\frac{p23\ r3}{-1+r1+r2}\right)$

*In[ ]:=* `{{p11,` $\frac{r1}{2}$ `/. ex}, {p12,` $\frac{r2}{3}$ `/. ex}, {p22,` $\frac{r2}{3}$ `/. ex}, {p23,` $\frac{r3}{2}$ `/. ex}}`

*Out[ ]=* `{{p11, 0.2}, {p12, 0.133333}, {p22, 0.133333}, {p23, 0.1}}`

Note that we cannot find analytical solutions, but the numerical solutions satisfy that the optimal design follows a balanced design in periods 1 and 3.

### Optimisation (approach 2) - assume balanced designs in periods 1 and 3

*In[ ]:=* `constr = term1 - term2;`

*In[ ]:=* `e1 = FullSimplify[Solve[D[term1, p12] == l D[constr, p12], l]]`
    `e2 = FullSimplify[Solve[D[term1, p22] == l D[constr, p22], l]]`
    `e3 = e1[[1]][[1]][[2]] == e2[[1]][[1]][[2]]`

*Out[ ]=* $\left\{\left\{l \rightarrow \left((-1+2\ p12+p22)\ (r1-4\ (-1+p12)\ p12\ r2)^2\right)\ \middle/ \right.\right.$
    $\left((-1+2\ p12+p22)\ r1^2-8\ p12\ (2\ p12^2+p12\ (-3+p22)\ -\ (-1+p22)\ (1+p22^2))\ r1\ r2\ +\right.$
    $\left.\left.\left.16\ p12^2\ (-1+p12+p22)\ (1+2\ p12^2+p22^2-p12\ (3+p22))\ r2^2\right)\right\}\right\}$

*Out[ ]=* $\left\{\left\{l \rightarrow \left(p12^2\ (-r1+4\ (-1+p12)\ p12\ r2)\right)\ \middle/ \left(-p12^2\ r1+(-1+p22)^2\ (-1+2\ p22)\ r1\ +\right.\right.\right.$
    $\left.\left.\left.4\ p12\ (-1+p12+p22)\ (1+p12^2-(3+p12)\ p22+2\ p22^2)\ r2\right)\right\}\right\}$

*Out[ ]=* $\left((-1+2\ p12+p22)\ (r1-4\ (-1+p12)\ p12\ r2)^2\right)\ \middle/$
    $\left((-1+2\ p12+p22)\ r1^2-8\ p12\ (2\ p12^2+p12\ (-3+p22)\ -\ (-1+p22)\ (1+p22^2))\ r1\ r2\ +\right.$
    $\left.16\ p12^2\ (-1+p12+p22)\ (1+2\ p12^2+p22^2-p12\ (3+p22))\ r2^2\right)\ ==$
    $\left(p12^2\ (-r1+4\ (-1+p12)\ p12\ r2)\right)\ \middle/\ \left(-p12^2\ r1+(-1+p22)^2\ (-1+2\ p22)\ r1\ +\right.$
    $\left.4\ p12\ (-1+p12+p22)\ (1+p12^2-(3+p12)\ p22+2\ p22^2)\ r2\right)$

*In[ ]:=* `sol2 = Solve[e3, {p12}];`

*In[ ]:=* **solsim = Simplify[sol2[[7]]]**

*Out[ ]=* $\Big\{ p12 \to$

$\dfrac{1}{24\,(-1+p22)\,r2}\Big(-4\,(3-6\,p22+2\,p22^2)\,r2 + \big(2\times 2^{1/3}\,(1-i\,\sqrt{3})\,r2\,\big((3-9\,p22+6\,p22^2)\,r1 +$

$(3-12\,p22+18\,p22^2-12\,p22^3+4\,p22^4)\,r2\big)\big)\Big/\Big(-9\,p22^2\,r1\,r2^2+27\,p22^3\,r1\,r2^2 -$

$18\,p22^4\,r1\,r2^2+18\,p22^2\,r2^3-72\,p22^3\,r2^3+108\,p22^4\,r2^3-72\,p22^5\,r2^3+16\,p22^6\,r2^3 +$

$\sqrt{\big(r2^3\,\big(-4\,\big((3-9\,p22+6\,p22^2)\,r1 + (3-12\,p22+18\,p22^2-12\,p22^3+4\,p22^4)\,r2\big)^3 +$

$p22^4\,r2\,\big(9\,(1-3\,p22+2\,p22^2)\,r1 -$

$2\,(9-36\,p22+54\,p22^2-36\,p22^3+8\,p22^4)\,r2\big)^2\big)\big)}\,\Big)^{1/3} +$

$2^{2/3}\,(1+i\,\sqrt{3})\,\Big(-9\,p22^2\,r1\,r2^2+27\,p22^3\,r1\,r2^2-18\,p22^4\,r1\,r2^2+18\,p22^2\,r2^3 -$

$72\,p22^3\,r2^3+108\,p22^4\,r2^3-72\,p22^5\,r2^3+16\,p22^6\,r2^3 +$

$\sqrt{\big(r2^3\,\big(-4\,\big((3-9\,p22+6\,p22^2)\,r1 + (3-12\,p22+18\,p22^2-12\,p22^3+4\,p22^4)\,r2\big)^3 +$

$p22^4\,r2\,\big(9\,(1-3\,p22+2\,p22^2)\,r1 -$

$2\,(9-36\,p22+54\,p22^2-36\,p22^3+8\,p22^4)\,r2\big)^2\big)\big)}\,\Big)^{1/3}\Big)\Big\}$

*In[ ]:=* **\$Assumptions = p12 > 0 && p22 > 0 && Element[p12, Reals] && Element[p22, Reals]**

*Out[ ]=* $p12 > 0\ \&\&\ p22 > 0\ \&\&\ p12 \in \mathbb{R}\ \&\&\ p22 \in \mathbb{R}$

*In[ ]:=* **Re[solsim]**

*Out[ ]=* $\Big\{ \mathrm{Re}\Big[ p12 \to$

$\dfrac{1}{24\,(-1+p22)\,r2}\Big(-4\,(3-6\,p22+2\,p22^2)\,r2 + \big(2\times 2^{1/3}\,(1-i\,\sqrt{3})\,r2\,\big((3-9\,p22+6\,p22^2)\,r1 +$

$(3-12\,p22+18\,p22^2-12\,p22^3+4\,p22^4)\,r2\big)\big)\Big/\Big(-9\,p22^2\,r1\,r2^2+27\,p22^3\,r1\,r2^2 -$

$18\,p22^4\,r1\,r2^2+18\,p22^2\,r2^3-72\,p22^3\,r2^3+108\,p22^4\,r2^3-72\,p22^5\,r2^3+16\,p22^6\,r2^3 +$

$\sqrt{\big(r2^3\,\big(-4\,\big((3-9\,p22+6\,p22^2)\,r1 + (3-12\,p22+18\,p22^2-12\,p22^3+4\,p22^4)\,r2\big)^3 +$

$p22^4\,r2\,\big(9\,(1-3\,p22+2\,p22^2)\,r1 -$

$2\,(9-36\,p22+54\,p22^2-36\,p22^3+8\,p22^4)\,r2\big)^2\big)\big)}\,\Big)^{1/3} +$

$2^{2/3}\,(1+i\,\sqrt{3})\,\Big(-9\,p22^2\,r1\,r2^2+27\,p22^3\,r1\,r2^2-18\,p22^4\,r1\,r2^2+18\,p22^2\,r2^3 -$

$72\,p22^3\,r2^3+108\,p22^4\,r2^3-72\,p22^5\,r2^3+16\,p22^6\,r2^3 +$

$\sqrt{\big(r2^3\,\big(-4\,\big((3-9\,p22+6\,p22^2)\,r1 + (3-12\,p22+18\,p22^2-12\,p22^3+4\,p22^4)\,r2\big)^3 +$

$p22^4\,r2\,\big(9\,(1-3\,p22+2\,p22^2)\,r1 -$

$2\,(9-36\,p22+54\,p22^2-36\,p22^3+8\,p22^4)\,r2\big)^2\big)\big)}\,\Big)^{1/3}\Big)\Big]\Big\}$

*In[ ]:=* **sol2 /. ex /. p22 → 0.23 / 0.8**

*Out[ ]=* $\big\{\{p12 \to -0.207107\},\ \{p12 \to 1.20711\},\ \{p12 \to -0.196053\},$

$\{p12 \to 0.908553\},\ \{p12 \to 0.866963 - 5.55112\times10^{-17}\,i\},$

$\{p12 \to -0.185879 - 2.77556\times10^{-17}\,i\},\ \{p12 \to 0.329662 + 1.11022\times10^{-16}\,i\}\big\}$

*In[ ]:=* **eq = FullSimplify[term1 - term2]**

*Out[ ]=* $\frac{1}{4}\left(r1 + \frac{4\,p12\,(-1 + p12 + p22)\,r2}{-1 + p22} + \right.$

$\frac{1}{r1 - 4\,(-1 + p12)\,p12\,r2}\left(r1^2 + r1\left(-1 + \left(-4\,(-1 + p12)\,p12 + (1 - 2\,p22)^2\right)r2\right) + \right.$

$\left.\left. 4\,p12\,r2\,(-1 + p12 + r2 + 4\,(-1 + p22)\,p22\,r2 + p12\,(-1 + 4\,p22)\,r2)\right)\right)$

*In[ ]:=* **eq3 = FullSimplify[e3]**

*Out[ ]=* $\left(r1 - 4\,(-1 + p12)\,p12\,r2\right)\left(p12^2\,/\,\left(-p12^2\,r1 + (-1 + p22)^2\,(-1 + 2\,p22)\,r1 + \right.\right.$

$4\,p12\,(-1 + p12 + p22)\left(1 + p12^2 - (3 + p12)\,p22 + 2\,p22^2\right)r2\right) +$

$((-1 + 2\,p12 + p22)\,(r1 - 4\,(-1 + p12)\,p12\,r2))\,/$

$\left((-1 + 2\,p12 + p22)\,r1^2 - 8\,p12\left(2\,p12^2 + p12\,(-3 + p22) - (-1 + p22)\left(1 + p22^2\right)\right)r1\,r2 + \right.$

$\left.\left.16\,p12^2\,(-1 + p12 + p22)\left(1 + 2\,p12^2 + p22^2 - p12\,(3 + p22)\right)r2^2\right)\right) = 0$

*In[ ]:=* **NSolve[{eq == 0 /. ex, eq3 /. ex}, {p12, p22}]**

*Out[ ]=* $\left\{\{p22 \to -28\,924.1 + 18\,543.2\,\text{i},\ p12 \to -0.249998 + 9.81706 \times 10^{-7}\,\text{i}\},\right.$

$\{p22 \to -0.0311394 + 0.265623\,\text{i},\ p12 \to -0.211592 + 0.0051594\,\text{i}\},$

$\{p22 \to -0.0311394 - 0.265623\,\text{i},\ p12 \to -0.211592 - 0.0051594\,\text{i}\},$

$\{p22 \to -0.0311394 + 0.265623\,\text{i},\ p12 \to -0.211592 + 0.0051594\,\text{i}\},$

$\{p22 \to 0.457912,\ p12 \to 0.153829\},\ \{p22 \to 0.457912,\ p12 \to 0.153829\},$

$\{p22 \to 0.7153 + 0.205041\,\text{i},\ p12 \to 0.151828 - 0.155807\,\text{i}\},$

$\{p22 \to 0.7153 - 0.205041\,\text{i},\ p12 \to 0.151828 + 0.155807\,\text{i}\},$

$\{p22 \to 1.84182,\ p12 \to 1.94225\},\ \{p22 \to 0.5,\ p12 \to 0.5\},\ \{p22 \to 0.5,\ p12 \to 0.5\},$

$\{p22 \to 0.117486 - 0.930047\,\text{i},\ p12 \to 1.17003 + 0.720353\,\text{i}\},$

$\{p22 \to 0.117486 + 0.930047\,\text{i},\ p12 \to 1.17003 - 0.720353\,\text{i}\},$

$\left.\{p22 \to 1.59697,\ p12 \to -0.816608\}\right\}$



\clearpage

\end{document}